\documentclass{PoS}

\usepackage{graphics,cite,amssymb,epsfig,float,psfrag}
\newcommand{\lsim}{\raisebox{-0.13cm}{~\shortstack{$<$ \\[-0.07cm] $\sim$}}~} 
\newcommand{\gsim}{\raisebox{-0.13cm}{~\shortstack{$>$ \\[-0.07cm] $\sim$}}~} 
\newcommand{\beq}{\begin{eqnarray}} 
\newcommand{\eeq}{\end{eqnarray}} 
\newcommand{\tb}{\tan \beta}

\title{Higgs Physics}

\ShortTitle{Higgs Physics}

\author{\speaker{Abdelhak Djouadi}\\
Laboratoire de Physique Th\'eorique, CNRS \& Universit\'e Paris--Sud, 91405 Orsay, France  \\
        E-mail: \email{abdelhak.djouadi@th.u-psud.fr}}

\abstract{
The implications of the discovery of a scalar Higgs boson at the LHC with a 
mass of approximately 125 GeV are summarised in the context of the Standard 
Model of particle physics with its unique scalar boson and of its most celebrated 
new physics extension, the minimal supersymmetric Standard Model or MSSM, in which 
the Higgs sector is extended to contain three neutral and two charged scalar bosons.  
Discussed are the implications from the measured mass, the production and decay 
rates of the observed particle and, in the MSSM, from the constraints in the 
search for the heavier Higgs states. The perspectives for Higgs
and new physics searches at the next LHC upgrades as well as at future hadron 
and lepton colliders are then briefly summarized.}

\FullConference{Proceedings of the Corfu Summer Institute 2014 "School and Workshops on Elementary Particle Physics and Gravity",\\
		3-21 September 2014\\
		Corfu, Greece }

\begin{document}

\section{Introduction}

The ATLAS and CMS historical discovery of a particle with a mass of  
125 GeV \cite{discovery} and pro\-per\-ties that are compatible with 
those of a scalar Higgs boson \cite{Higgs,Review1} has far reaching 
consequences not only for the Standard Model (SM) but also for new 
physics models beyond it. In the SM, electroweak symmetry  breaking is 
achieved spontaneously via the Brout--Englert--Higgs mechanism \cite{Higgs}, 
wherein the neutral  component of an isodoublet scalar field acquires a 
non--zero vacuum  expectation value $v$. This gives rise to nonzero masses for 
the fermions and  the electroweak gauge bosons while preserving the  
${\rm SU(2)\!\times\! U(1)}$ gauge symmetry. One of the four degrees of 
freedom of the original isodoublet field, corresponds to a physical particle
\cite{Review1}: a scalar  boson with  ${\rm J^{PC}}\!=\!0^{++}$ quantum numbers 
under parity and charge conjugation. The couplings of the Higgs boson to the 
fermions and gauge bosons are  related to the  masses of these particles and 
are thus decided by the symmetry breaking mechanism. In contrast, the mass of 
the Higgs boson itself $M_H$, although expected to be in the vicinity of the 
weak scale $v\!\approx\! 250$ GeV,  is undetermined.  Let us summarise the 
known information on this parameter before the start of the LHC.

A direct information  was the lower limit  $M_H \! \gsim 114$ GeV at 95\% 
confidence level (CL) established at LEP2 \cite{PDG}. Furthermore, a global 
fit of the electroweak precision data to which the Higgs boson contributes, 
yields the value $M_H=92^{+34}_{-26}$ GeV,
corresponding to a 95\% CL upper limit of $M_H \lsim 160$ GeV \cite{PDG}.
From the theoretical side, the presence of this new weakly coupled degree of
freedom is a crucial ingredient  for a unitary electroweak theory. Indeed, the
SM without the Higgs particle leads  to scattering amplitudes of the $W/Z$  
bosons that grow with the square of the center of mass energy and perturbative 
unitarity would be lost at energies above the  TeV scale.  In fact, even in the 
presence of a Higgs boson, the $W/Z$ bosons could  interact very strongly with each
other and, imposing the unitarity requirement in the high--energy scattering 
amplitudes leads to the important mass bound $M_H \lsim 700$ GeV \cite{H-LQT}, 
implying that the particle is kinematically accessible at the LHC. 

Another theoretical constraint emerges from the fact that the Higgs self--coupling, 
$\lambda \propto M_H^2$, evolves with energy and at some stage, becomes very large 
and even infinite and the theory completely looses its predictability. If the energy scale 
up to which the couplings remains finite is of the order of $M_H$ itself, one should 
have $M_H \lsim$ 650 GeV \cite{triviality}. On the other hand, for small values of 
$\lambda$ and hence $M_H$, the quantum corrections tend to drive the self--coupling to 
negative values  and completely destabilize the scalar Higgs potential to the point where 
the minimum is not stable anymore \cite{triviality}. Requiring $\lambda \geq 0$,  
up to the TeV scale implies 
that $M_H \gsim 70$ GeV. If the SM is to be extended to the Planck scale ${\rm 
M_P \sim 10^ {18}}$ GeV, the requirements on $\lambda$ from finiteness and positivity 
constrain the Higgs mass to lie in the range 130 GeV$\lsim M_H \lsim$ 180 GeV 
\cite{triviality}. This  narrow margin is close
to the one obtained from  the direct and indirect experimental constraints. 

The discovery of the Higgs particle with a mass of 125 GeV, a value that makes the
SM perturbative, unitary and extrapolable to the highest possible scales,  
is therefore a consecration of the model and crowns its past success in describing 
all experimental data available. In particular, the average mass value measured by 
the ATLAS and CMS teams, $M_H \! = \! 125.1 \pm 0.24$ GeV \cite{LHC-update}, 
is remarkably close to the best--fit of the precision data which should be considered 
as a great achievement and a triumph for the SM. In addition, a recent analysis that
includes the state-of-the-art  quantum corrections \cite{stability} 
gives for the condition of absolute 
stability of the electroweak vacuum, $\lambda (M_P) \geq 0$,  the bound $M_H \gsim 
129$ GeV for the present value of the top quark mass and the strong coupling constant,
$m_t^{\rm exp}=173.2 \pm 0.9~{\rm GeV}$ and $\alpha_s(M_Z)=0.1184 \pm 0.0007$ \cite{PDG}. 
Allowing for a 2$\sigma$ variation of $m_t^{\rm exp}$, one obtains $M_H \! \geq \! 125.6$ GeV 
that is close to the measured $M_H$ value \cite{LHC-update}. In fact, for an unambiguous and  
well-defined determination of the top mass,  one should rather use the total cross
section for top pair production at hadron colliders which can unambiguously be defined 
theoretically; this mass has a larger uncertainty, $\Delta m_t \approx 3$ GeV, which 
allows  more easily for the absolute
stability of the SM vacuum up to $M_P$; see Ref.~\cite{stability0} for discussions
on this issue. 

Nevertheless, the SM is far from being perfect in many respects. 
It does not explain  the proliferation of fermions and the large hierarchy 
in their  mass spectra and does not say much about the small neutrino masses. 
The SM does not unify in a satisfactory way the electromagnetic, weak and strong  
forces, as one has three different symmetry groups with three coupling 
constants which shortly fail to meet at a common value during their evolution with 
the energy scale;   it also ignores the fourth force, gravitation. Furthermore, it 
does not contain a  particle that could account for the cosmological dark matter 
and fails to explain the baryon asymmetry in the Universe.

However, the main problem that calls for beyond the SM is related to the special
status of the Higgs boson which, contrary to fermions and gauge bosons has a mass 
that cannot be protected against quantum corrections. Indeed, these are quadratic in the 
new physics scale which serves as a cut--off and hence, tend to drive $M_H$ to very 
large values, ultimately to $M_P$, while we need $M_H = {\cal O}(100~{\rm GeV})$.
Thus, the SM cannot be extrapolated beyond ${\cal O}(1~$ TeV) where new physics 
should emerge.  This is the reason why we expect something 
new to manifest itself at the LHC.  

There are three avenues for the many possibilities for this new physics beyond the SM 
\cite{BSM}.
There are first theories with extra space--time dimensions that emerge at the 
TeV scale (the cut--off is then not very high) and, second, composite models inspired from 
the strong  interactions also at the TeV scale (and thus the Higgs is not a fundamental spin--zero
particle). Some versions of these scenarios do not incorporate any Higgs particle in their 
spectrum and are thus ruled out by the Higgs discovery. However, the option that emerges 
in the most natural way is Supersymmetry (SUSY) \cite{SUSY} as it solves most of the
SM problems discussed above. In particular,  SUSY protects $M_H$ as the quadratically 
divergent radiative corrections from standard particles are exactly compensated by the 
contributions of their supersymmetric partners. These sparticles should  not be
much heavier 
than 1 TeV not to spoil this compensation \cite{natural} and they should be  
thus produced at the LHC.

The Higgs discovery is very important for SUSY and, in particular, for its 
simplest low energy manifestation, the minimal supersymmetric Standard Model 
(MSSM) that indeed predicts a light Higgs state. In 
the MSSM, two Higgs doublet  fields $H_u$ and $H_d$ are required, leading to an 
extended Higgs consisting of five Higgs bosons, two CP--even $h$ and 
$H$, a CP--odd $A$ and two charged $H^\pm$ states \cite{Review2}. Nevertheless,
 only two parameters are needed to describe the Higgs sector at tree--level: one Higgs mass, 
which is generally taken to be that of the pseudoscalar $A$ boson $M_A$, and the 
ratio of vacuum expectation values of the two Higgs fields, $\tan\beta =v_d/v_u$, 
expected to lie in the range $1\! \lsim \! \tb \! \lsim \! 60$. The masses of 
the CP--even $h,H$ and the charged $H^\pm$ states, as well as the mixing angle 
$\alpha$ in the CP--even sector are uniquely defined in terms of 
these two inputs at tree-level, but this nice property is spoiled at higher orders
\cite{CR-Higgs}. For $M_A \! \gg \! M_Z$, one is in the so--called decoupling
regime in which the $h$ state is light and has almost exactly the SM--Higgs couplings, 
while the  other CP--even $H$ and the charged $H^\pm$ bosons become heavy, $M_H \! \approx
\! M_{H^\pm} \! \approx \! M_A$, and decouple from the massive gauge bosons. In this regime, 
the MSSM Higgs sector thus looks almost exactly as the one of  the SM with its unique 
Higgs boson.

Nevertheless, contrary to the SM Higgs boson, the lightest MSSM CP--even $h$ mass is 
bounded from above and, depending on the SUSY parameters that enter the important quantum
corrections, is restricted to $M_h^{\rm max} \lsim$ 130 GeV \cite{CR-Higgs} if one assumes
a SUSY breaking scale that is not too high, $M_S\! \lsim\! {\cal O}$ (1 TeV), in order
to avoid too much fine-tuning in the model.  Hence, the requirement that the 
MSSM $h$ boson coincides with the one observed at the LHC, i.e. with $M_h \approx 125$ 
GeV and almost SM--like couplings as the LHC data seem to indicate, would place very 
strong constraints on the MSSM parameters, in particular the SUSY--breaking scale $M_S$.  
This comes in addition to the limits obtained from the search of the heavier 
Higgs states and the superparticles at the LHC. 

In this lecture, the implications of the discovery of the Higgs boson at the LHC  and the 
measurement of its couplings to other particles will be summarised and the 
prospects for future searches of new physics, in particular in the SUSY context,  
will be summarized.\vspace*{-2mm} 
 

\section{Implications for the Standard Model}
\vspace*{-2mm} 

In many respects, the Higgs particle was born under a very lucky star as the 
mass value of $\approx 125$ GeV allows 
to produce it at the  LHC in many redundant channels  and to detect it in a  
variety of decay modes. This allows detailed studies of the Higgs properties
as will be discussed here.

\subsection{Higgs production and decay}

\begin{figure}[hbtp]
\begin{center}
\vspace*{.1cm}
\hspace*{-1.5cm}
\epsfig{file=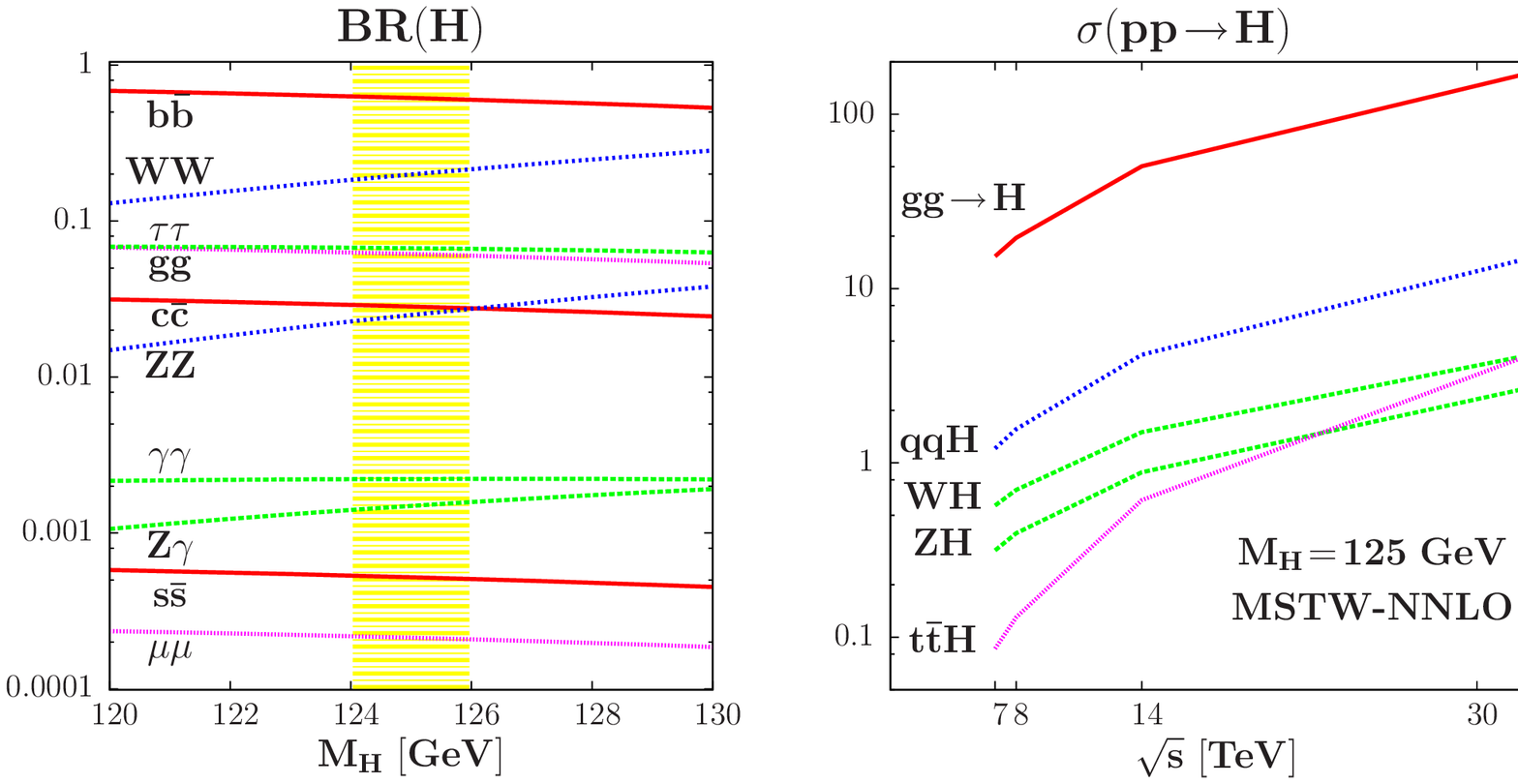,width=13.2cm}
\vspace*{-.8cm}
\end{center}
\caption{The SM--like Higgs boson branching ratios in the mass range $120$--130
GeV (left) and its production cross sections at proton colliders as a function of  
the c.m. energy (right); from Ref.~\cite{ratios}.}
\label{fig:allH}
\vspace*{-.4cm}
\end{figure}

We start by summarizing the production and decay at the LHC of a light SM--like 
Higgs particle, which should correspond to the lightest MSSM $h$ boson in the 
decoupling regime.  First, for $M_H \approx 125$ GeV, the Higgs mainly decays 
into $b \bar b$ pairs but the decays into $WW^*$ and $ZZ^*$ final states, before 
allowing the gauge bosons to decay leptonically $W \! \to \! \ell \nu$ and $Z\! \to 
\! \ell \ell$ ($\ell\! =\! e,\mu$), are also significant. The $H\! \to \! \tau^+\tau^-$ 
channel (as well as the $gg$ and $c\bar c$ decays that are not detectable at the LHC) 
is also of significance,  while the clean loop induced $H\to \gamma \gamma$ mode can 
be easily detected albeit its small rates. The very rare $H\to Z\gamma$  and even 
$H\to \mu^+\mu^-$ channels should be accessible at the LHC but only with a much larger 
data sample. This is illustrated in Fig.~\ref{fig:allH} (left) where the Higgs decay 
branching fractions are displayed for the narrow mass range $M_H=120$--130 GeV 
\cite{HDECAY}. 
 
On the other hand, many Higgs production processes have significant cross sections 
\cite{Baglio,LHCXS} as is shown in the right--hand side of Fig.~\ref{fig:allH} where they 
are displayed at a proton collider at various past, present and foreseen center of mass 
energies for a 125 GeV SM--like Higgs boson. While the by far dominant gluon 
fusion mechanism $gg\to H$  (ggF) has extremely large rates ($\approx\! 20$ pb at 
$\sqrt s\!=\! 7$--8 TeV), the subleading channels, i.e. the vector boson fusion (VBF) 
$qq \to V^* V^* \to Hqq$ and the Higgs--strahlung (HV) $q\bar q \to HV$ with $V=W,Z$ mechanisms, 
have cross sections which should allow for a study of the Higgs particle already at $\sqrt 
s\gsim 7$ TeV with the amount of integrated luminosity, $\approx 25$ fb$^{-1}$, that has 
been collected by each experiment. The Higgs--top associated process $p p\to t\bar 
t H$ (ttH) would require higher energy and luminosity. 

This pattern already allows ATLAS and CMS to observe the Higgs boson
in several channels and to measure some of its couplings in a reasonably accurate way. 
The channels that have been searched are $H \! \to \! ZZ^* \! \to \! 4\ell^\pm,  
H\! \to \! WW^* \! \to \! 2\ell 2 \nu, H \! \to \! \gamma\gamma$ where the Higgs is 
mainly produced in ggF with subleading contributions from $Hjj$  
in the VBF process, $H \! \to \! \tau \tau$  where the Higgs is produced in 
association with one (in ggF) and two (in VBF) jets, and finally $H\to b \bar b$ 
with the Higgs produced in the HV process. One can ignore for the moment
the additional search channels $H\! \to \! \mu \mu$ and $H \! \to \! Z\gamma$
for which the sensitivity is still too low with the  data collected so far.

\begin{figure}[!h]
\vspace*{-1mm}
\begin{center}
\begin{tabular}{ll}
\begin{minipage}{4cm}
\hspace*{-1cm}
\epsfig{file=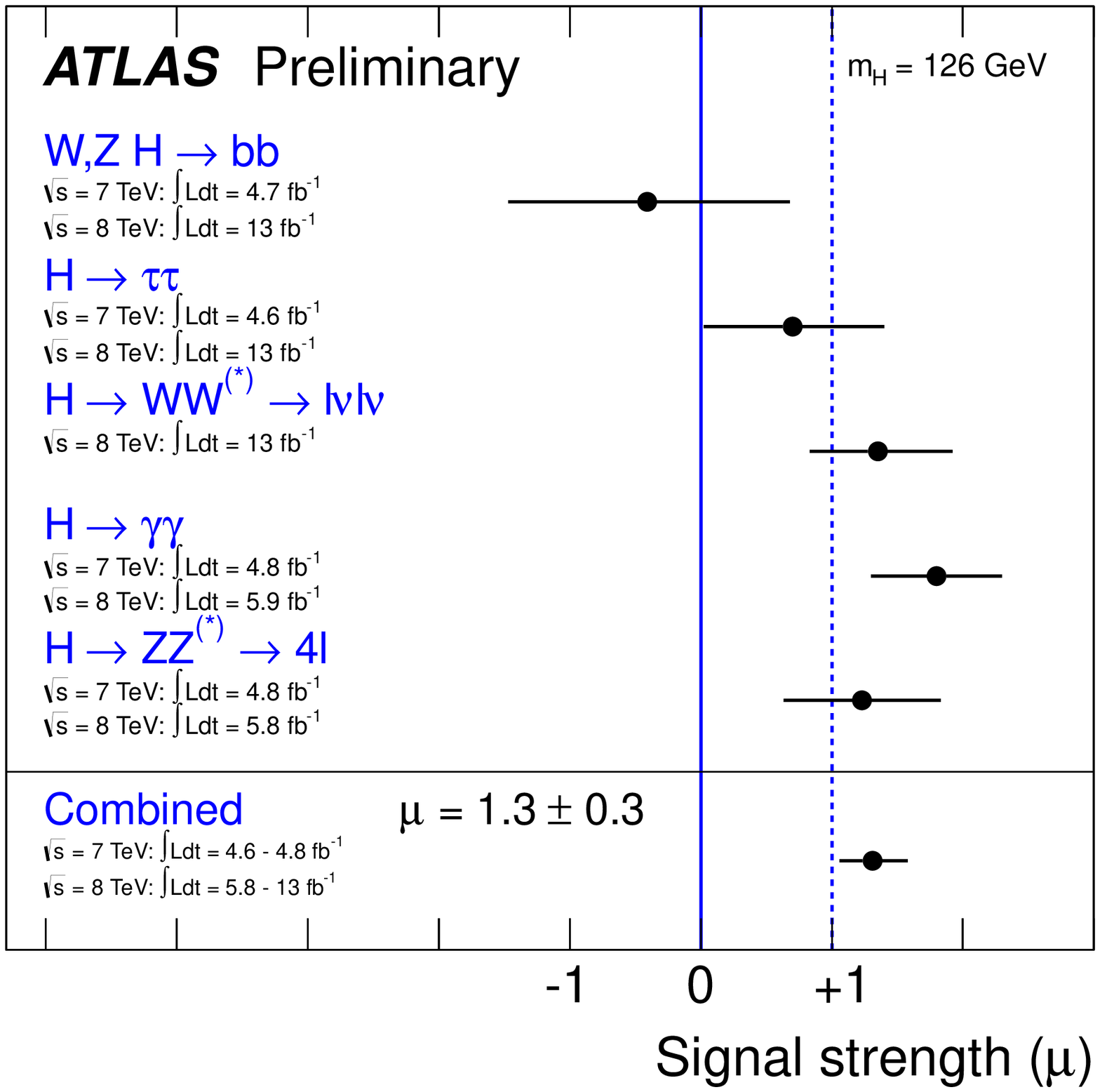,width=5.9cm}\hspace*{.3cm}
\end{minipage}
& \hspace*{2cm} 
\begin{minipage}{4cm}
\epsfig{file=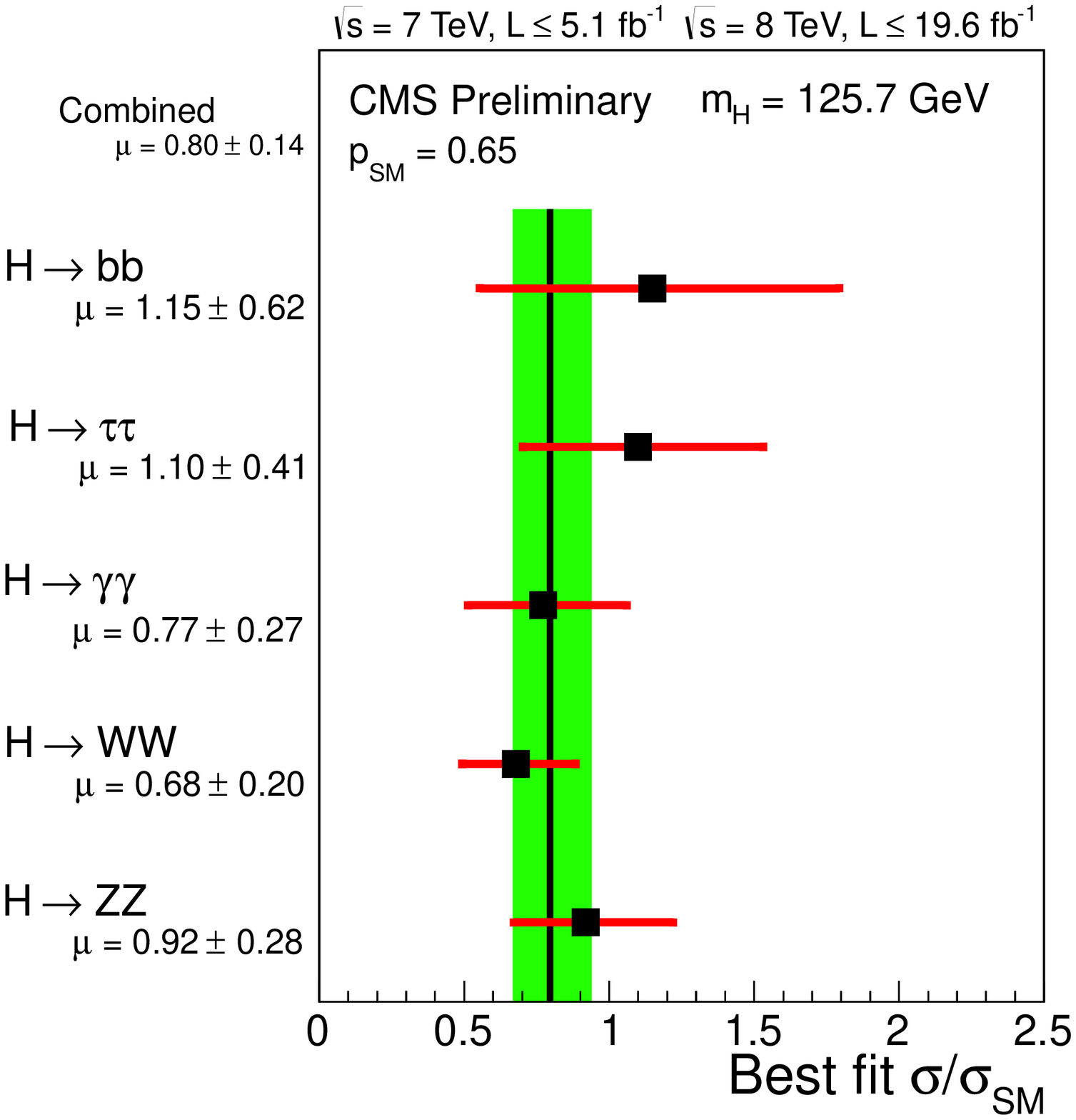,width=5.9cm}\hspace*{-.3cm}
\end{minipage}
\end{tabular}
\end{center}
\vspace*{-6mm}
\caption{The signal strengths on the SM Higgs boson in the various search channels provided 
by ATLAS and CMS with the data collected so far at $\sqrt s=7$+8 TeV; from the summary given in
Ref.~ \cite{LHC-update}.
}
\label{Fig:constraints}
\vspace*{-2mm}
\end{figure}

A convenient way to scrutinize the couplings of the produced $H$ boson is to look at
their deviation from the SM expectation. One then considers for a given search 
channel the signal strength modifier $\mu$ which for the $H\! \to \! XX$ decay mode
measures the deviation compared to the SM expectation 
of the Higgs production cross section times  decay branching fraction
\beq
\mu_{XX}\vert_{\rm th} & = & \frac {\sigma( pp \to H \to XX)}{ \sigma( pp \to H
\to XX)|_{\rm SM}} =   \frac {\sigma( pp \to H)\times {\rm BR} (H \to XX)}{
\sigma( pp \to H)|_{\rm SM} \times {\rm BR} (H \to XX)|_{\rm SM} } .
\label{mudef}  
\eeq  
ATLAS and CMS have provided the signal strengths for the various final states
with a luminosity of $\approx 5$ fb$^{-1}$ for  the 2011 run at $\sqrt s=7$ TeV 
and $\approx 20$ fb$^{-1}$ for the 2012 run at $\sqrt s=8$ TeV.  The constraints 
given by the two collaborations are shown in Fig.~\ref{Fig:constraints}. 
When combined, this leads to a global signal strength \cite{LHC-update}: 
\begin{eqnarray}  
\mu_{\rm tot}^{\rm ATLAS} =1.18 \pm 0.15\ , \ \ 
\mu_{\rm tot}^{\rm CMS}  =1.00 \pm 0.14   \label{muvalues} 
\end{eqnarray}
The global signal strength being very close 
to unity, it implies that the observed Higgs is  SM--like.

\subsection{The theoretical uncertainties and the signal strength ratios}\vspace*{-1mm} 

As can be seen, already  with the rather limited statistics at hand, the accuracy of the 
ATLAS and CMS measurements  in eq.~(\ref{muvalues}) is reaching  the 15\% level. This is  
at the same time impressive and worrisome. Indeed, the main Higgs production channel is 
the top and bottom quark loop mediated gluon fusion mechanism and, at $\sqrt s\!=\!7$ or 8 TeV, the  three other mechanisms contribute at a total level below 15\%.  
The majority of the signal events observed at LHC, in particular in the search channels  
$H \! \to \!  \gamma \gamma, H \! 
\to \!   ZZ^*  \to  4\ell,  H  \! \to \!  WW^*  \to 2 \ell 2\nu$
and, to a lesser extent $H \!   \to \!   \tau\tau$, thus come
from the ggF mechanism which  is known to be affected by large
theoretical uncertainties.

Indeed, although  $\sigma(gg\to H)$ is known up 
next--to--next--to--leading order (NNLO) in perturbative QCD 
(and at least at NLO for the electroweak interaction) \cite{HO-QCD,HO-EW,Baglio,LHCXS}, there is a
significant  residual scale dependence which points to the possibility that
still higher order contributions cannot be  totally excluded. In
addition, as the process is of ${\cal O}(\alpha_s^2)$ at LO and is initiated by
gluons, there are sizable uncertainties due to the  gluon parton distribution
function (PDF) and the value of the coupling $\alpha_s$. A third source of 
theoretical uncertainties, the use of an effective field theory (EFT) approach 
to calculate the radiative corrections  beyond the NLO approximation, should in 
principle also be considered \cite{Baglio}. In addition, large uncertainties 
arise when the $gg\! \to\! H$ cross section is broken into the jet categories
$H\! +\! 0j, H\! +\! 1j$ and $ H\! +\! 2j$ \cite{LHCXS2}. In total, 
the combined theoretical uncertainty is estimated to be $\Delta^{
\rm th} \! \approx \! \pm 15\%$ \cite{LHCXS} and would  increase to 
$\Delta^{\rm th}   \! \approx \pm $20\% if the EFT uncertainty is also 
included. The  a priori cleaner VBF process will be contaminated by the  $gg\! \to\! 
H\!+\!2j$ mode making the total  uncertainty in the $H\!+\!jj$ ``VBF" sample 
also rather large \cite{LHCXS2}. 

Hence, the theoretical uncertainty is  already at the level of the accuracy of 
the cross section measured by the ATLAS and CMS collaborations, eq.~(\ref{muvalues}).
Another drawback of the analyses
 is that they involve strong theoretical assumptions on the
total Higgs width since some contributing decay channels  not accessible at the
LHC are assumed  to be SM--like and possible invisible Higgs decays in scenarios
beyond the SM are supposed not to occur. In Ref.~\cite{ratios}, following earlier work 
\cite{Dieter}, it has been suggested to 
consider the decay ratios $D_{XX}$  defined as 
\beq
D_{XX}^{\rm p} \! = \! \frac {\sigma^{\rm p} ( pp \to H \to XX)}{ \sigma^{\rm p}
( pp \to H \to VV)} \! =\!   \frac {\sigma^{\rm p} ( pp \to H)\times {\rm BR} 
(H \to XX)}{ \sigma^{\rm p}( pp \to H) \times {\rm BR} (H \to VV)} \! =\!  
\frac{\Gamma(H \to XX)}{ \Gamma ( H \to VV)}  \propto \frac{c_{X}^2} {c_{V}^2} 
\label{ratio} 
\eeq 
for a specific production process $p$  and  for a given decay channel $H\to XX$ when 
the reference channel $H\to VV$ is used. In these ratios,  the cross sections $\sigma^p 
(pp \! \to \! H)$ and hence, their significant theoretical  uncertainties will cancel out, 
leaving out only the ratio of partial decay widths which are affected 
by much smaller uncertainties. The total decay
width which includes contributions from channels not under control such as
possible invisible Higgs decays, do  not appear in the ratios $D_{XX}^{\rm
p}$.  Some common experimental systematical uncertainties such as the one from
the luminosity measurement and the small uncertainties in the Higgs decay 
branching ratios also cancel out. We are thus left with only with the statistical 
and some (non common) systematical errors \cite{ratios}.

The ratios $D_{XX}$ involve, up to  kinematical factors and known radiative
corrections, only the ratios  $\vert c_X \vert^2/$ $\vert c_V\vert^2$ of the 
Higgs reduced  couplings to the particles  $X$ and $V$ compared to the SM expectation, 
$c_X \equiv g_{HXX}/g_{HXX}^{\rm SM}$. For the time being, three independent ratios 
can be considered:  $D_{\gamma \gamma}, D_{\tau \tau}$ and $D_{bb}$. In order to determine
these ratios,   the theoretical uncertainties have to be treated 
as a bias (and not as if they were associated with a statistical distribution) and 
the fit should be performed for the two $\mu$ extremal values: 
$\mu_{i} \vert_{ \rm exp} \pm \delta \mu_i/\mu_i \vert_{\rm th}$ with  
$\delta\mu_i/ \mu_i \vert_{\rm th} \approx \pm 20\%$ \cite{fit}.

\begin{figure}[!h] 
\vspace*{-3.6cm}
\begin{center}
\begin{tabular}{lll}
\vspace*{-3.5cm}
\begin{minipage}{15.5cm}
\vspace*{-5mm}
\epsfig{file=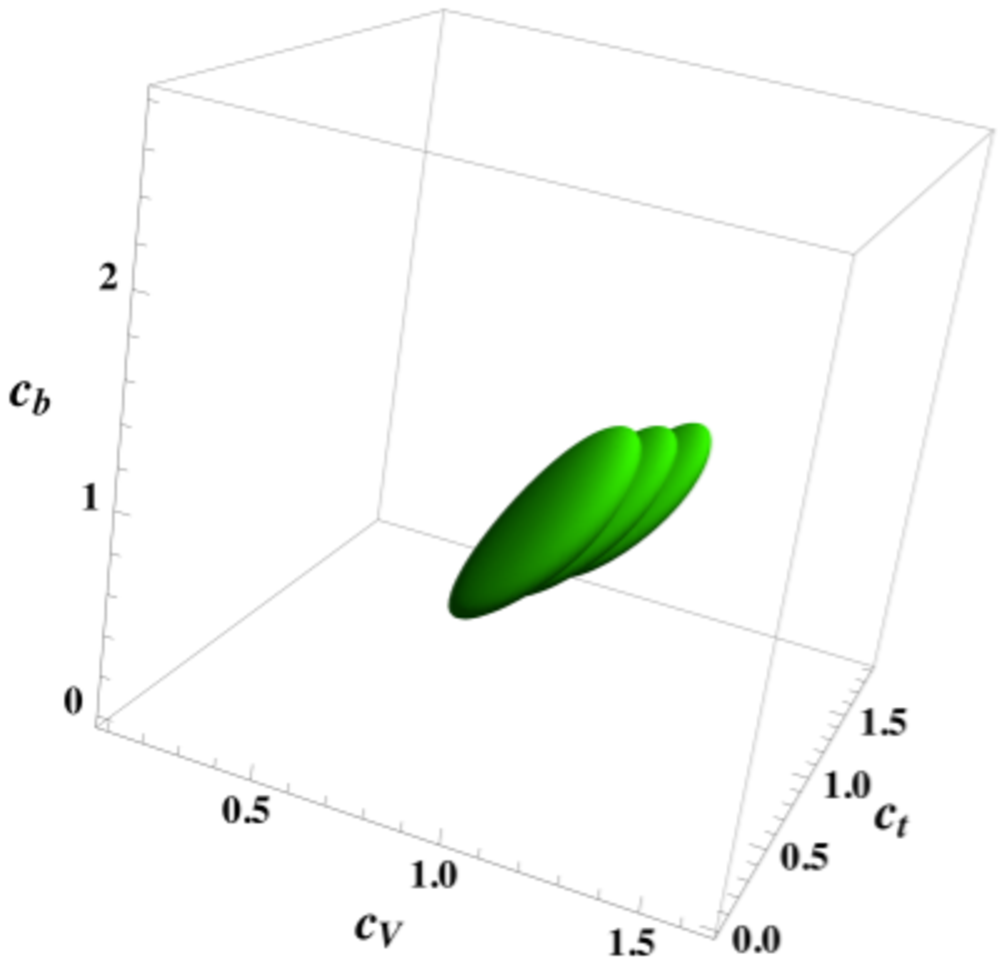,width=9cm}
\end{minipage}
& \hspace*{-7.1cm}
\begin{minipage}{10cm}
\epsfig{file=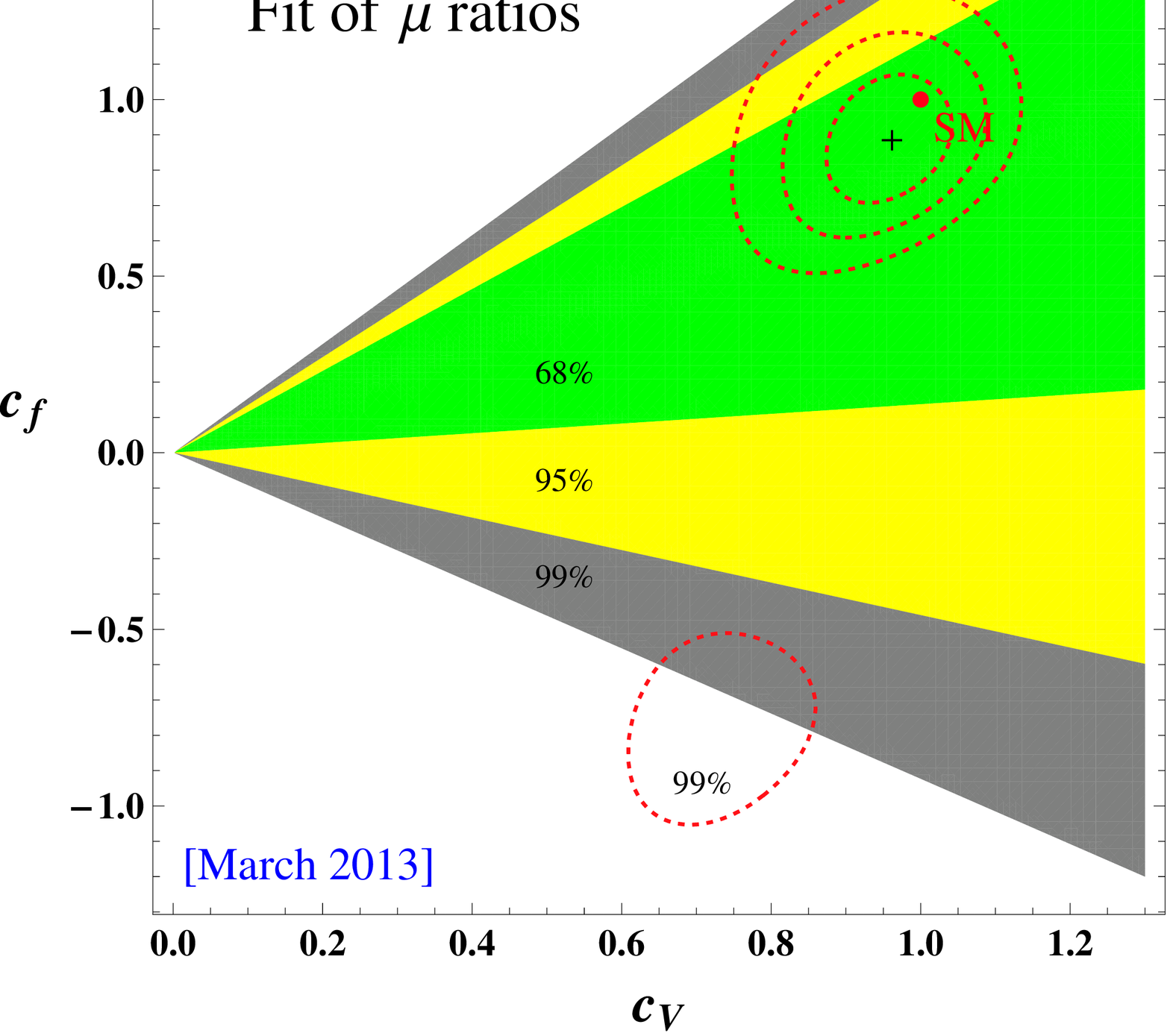,width=5cm}
\vspace*{-2.cm}
\end{minipage}
\end{tabular}
\end{center}
\vspace*{-.8cm}
\caption{{
Left:  
the best-fit region at $68\%{\rm CL}$  
for the Higgs signal strengths in the $[c_t,c_b,c_V]$ space \cite{Habemus}; the 
three overlapping regions are for the central and extreme  choices of 
the theoretical prediction for the Higgs rates including uncertainties.
Right: best-fit regions at $68\%{\rm CL}$ (green), $95\%{\rm
CL}$ (yellow) and $99\%{\rm CL}$ (grey) in the plane $c_f$ versus $c_V$;
the associated best-fit point (cross) and SM (red) point are also shown \cite{fit}.  
}}
\label{fig:3D}
\vspace*{-1mm}
\end{figure}

A large number of analyses of the Higgs couplings from the LHC data have 
been performed and in most cases, it is assumed that the 
couplings of the Higgs boson to the massive $W,Z$ gauge bosons are equal, 
$g_{HZZ}=g_{HWW}=c_V$, and the couplings to all fermions are also the 
same $g_{Hff}=c_f$. However, as for instance advocated in Ref.~\cite{Habemus}
to characterize the Higgs particle at the LHC, at least three independent $H$ 
couplings should be considered, namely $c_t$, $c_b$ and $c_V$. One can 
thus define the following effective Lagrangian (in which the custodial symmetry 
relation  $c_W=c_Z$ which holds in most cases is assumed) for the Higgs interactions
\begin{eqnarray} 
{\cal L}_H  \! =\!    c_V  g_{HVV}  H  V_{\mu} V^{\mu} +
  c_t y_t  H  \bar t_L  t_R \! -\!  c_t  y_c H  \bar c_L  c_R \! -\! 
  c_b  y_b   H  \bar b_L b_R  \! - \! c_b y_\tau H  \bar \tau_L \tau_R 
 \! + \! {\rm h.c.}
\end{eqnarray}
where $y_{t,c,b,\tau}=m_{t,c,b,\tau}/v$ are the Yukawa couplings of the heavy
SM fermions, $g_{HWW}\! = \! 2M^2_W/v$ and $g_{HZZ}\! =\! M^2_Z/v$ the $HWW$
and $HZZ$ couplings and $v$ the SM Higgs vev.  While the couplings to $W,Z,b,\tau$
particles are derived by considering the decays of the Higgs boson to these particles, 
the $Ht\bar t$ coupling is derived indirectly from $\sigma( gg\! \to\! H)$ 
and BR($H\! \to\! \gamma \gamma)$, two processes that are generated by triangular loops 
involving the top quarks in the SM. One can assume, in a first approximation, that 
$c_c=c_t$ and $c_\tau=c_b$ and possible invisible Higgs decays are absent. 
In Ref.~\cite{Habemus}, a three--dimensional fit of the $H$ couplings was performed 
in the space $[c_t, c_b, c_V]$, when the theory uncertainty is taken
as a bias and not as a nuisance.  The results of this fit are presented in 
Fig.~\ref{fig:3D} (left) for $c_t,c_b,c_V \! \geq \! 0$. The best-fit value for the 
couplings, with  the $\sqrt s=7+$8 TeV ATLAS and CMS data turns out to be   
$c_t=0.89, ~ c_b=1.01$ and $c_V=1.02$, ie very close to the SM values.

In the right-hand side of Fig.~\ref{fig:3D},  a universal Higgs coupling to 
fermions $c_t=c_b=c_f$ is assumed and shown are the results when fitting the 
signal strengths as well as the Higgs decay branching ratios in
the various modes. 
One sees that  the best-fit domains from the ratios obtained e.g. at $1\sigma$ do 
not exclude parts of the $1\sigma$ regions obtained from the $\chi^2$  since the main
theoretical uncertainty cancels out in the $D_{XX}$ ratios   and is negligible
for the signal strengths when added in quadrature to the experimental
error.  The domains from the likelihood fit are more restricted
as the fit  exploits the full experimental  information on the
Higgs rates and not only on the ratios and also, as the experimental  error
on a ratio of rates is higher than on the rates alone. The situation
might improve in the future when the experimental uncertainty will decrease. 

\subsection{Implications of the Higgs couplings measurement}\vspace*{-2mm}

\noindent {\bf a) A fourth generation fermions?} 

One of the immediate implications of the LHC 
Higgs results is that a fourth generation of fermions (with a heavy neutrino 
$m_{\nu'} \gsim \frac12 M_Z$ to evade le LEP bounds \cite{PDG}) is now completely 
ruled out \cite{4th:Lenz}. Indeed, in addition to the direct LHC searches that 
exclude heavier quarks $m_{b'},  m_{t'} \! \lsim \! 600$ GeV \cite{4th:LHC}, strong 
constraints can be also obtained from the loop induced Higgs--gluon and Higgs-photon 
vertices in which any heavy particle coupling to the Higgs proportionally to its mass
will contribute.  In particular, for the $gg\to H$ leading process, the 
additional 4th generation $t'$ and $b'$ contributions increase the rate 
by  a factor of $\approx 9$ at LO. However, there are large ${\cal O}(G_F
m_{f'}^2)$ electroweak corrections that affect these 
vertices, leading to a strong suppression of the  $gg\to H\!\to\! \gamma
\gamma$ rate, making this channel unobservable with four generations.
 Using a version of  {\tt HDECAY} with a fourth generation, one can show 
that the rate $\sigma(gg\! \to\! H)\!\times\! {\rm BR}(  H\! \to\! 
\gamma \gamma)|_{\rm SM4/SM}$
for $M_H\!=\!125$ GeV is a factor of 5 to 10 smaller than in the SM. Also,  
the ratio $\sigma(q\bar q\! \to\! VH)\!\times\! {\rm BR}(  H\! \to\! b\bar
b)|_{\rm SM4/SM}$  is reduced by a factor 3 to 5 depending on $m_{\nu'}$. 
Hence with a fourth family, the Higgs signal would have not been observable 
and the obtained Higgs results unambiguously rule out  this possibility.\smallskip

\noindent {\bf b) Total width and invisible Higgs decays}

Invisible decays would affect the properties of the observed Higgs particle and  
could be constrained if the total decay width is determined. But for a 125 GeV 
Higgs particle, $\Gamma_H^{\rm tot} =4$ MeV, is too small to be resolved experimentally.   
Nevertheless, in $pp \to VV  \to 4f$, a large fraction of the Higgs cross 
section lies in the high--mass tail \cite{width:Passarino+Kauer}. This allows to constrain  
$\Gamma_H^{\rm tot}$
and recent measurements of the ATLAS and CMS collaborations lead to a bound  $\Gamma_H^{\rm 
tot}/\Gamma_H^{\rm SM} \approx 5$--10 \cite{LHC-update}. These bounds are
thus too loose and in addition strongly rely on the assumption that the off--shell 
Higgs couplings are exactly the same as the on--shell couplings, which is not true in many cases. 
 
Nevertheless, the invisible Higgs decay width can be constrained indirectly by a fit 
of the Higgs couplings and in particular with the signal strength in the 
$H\! \to \! ZZ$ process: $\mu_{ZZ}\! \propto\! \Gamma (H\! \to \! ZZ)/\Gamma_H^{\rm tot}$ 
with $\Gamma_H^{\rm tot} \! = \! \Gamma_H^{\rm inv}\! + \! \Gamma_H^{\rm SM}$ and 
$\Gamma_H^{\rm SM}$ calculated  with free coefficients $c_f$ and $c_V$.  The measurements
give $\Gamma_H^{\rm inv}/  \Gamma_H^{\rm SM} \lsim 50\%$ at the $95\%~{\rm CL}$ 
if the assumption $c_f=c_V=1$ is made \cite{fit}. 

A more model independent approach would be to perform direct searches for missing
transverse energy. These have been conducted by ATLAS and CMS in the $pp\to HV$ process 
with $V\! \to \! jj, \ell \ell$ and in the VBF channel, $qq \to qq E_T\hspace*{-3mm}\slash$. 
For $M_H \! \approx \! 125$ GeV a bound BR$_{\rm inv} \lsim 50\%$ is obtained at the 
95\%CL if the Higgs couplings are SM--like \cite{LHC-update}. 
A more promising search for invisible decays is the monojet channel $gg\!\to\! Hj$ which 
has large rates  \cite{monojet}. Hence, if the Higgs is coupled to invisible particles, 
it may recoil against hard QCD radiation, leading to monojets. While the most recent monojet 
ATLAS and CMS searches \cite{LHC-update} are only sensitive to an invisible rate close to 
unity, more restrictive results can be obtained with more data at the upgraded LHC. 

The Higgs invisible rate and the dark matter detection rate in direct astrophysical 
searches are correlated in Higgs portal models. Considering the generic cases of 
scalar, fermionic and vectorial dark matter particles $\chi$ that couple only to the 
Higgs, one can translate in each case the LHC constraint  ${\rm BR} (H \to {\rm inv.})$ 
into a constraint on the Higgs couplings to the $\chi$ particles.  It turns out that
these constraints are competitive \cite{invis-portal} with those derived from the bounds on
the dark matter scattering cross section on nucleons \cite{DM-direct} 
from experiments such XENON for instance.\smallskip

\noindent {\bf c) Determination of the spin--parity quantum numbers}

Apart from the measurement of the couplings, one also needs  to
establish that the observed Higgs state is indeed a CP even scalar
and hence with ${\rm J^{PC}= 0^{++}}$ quantum numbers. For the spin, 
the observation of the $h\to \gamma\gamma$ decay rules out the 
spin--1 case by virtue of the Landau--Yang theorem \cite{Landau-Yang}. 
The Higgs parity 
can be probed by studying  kinematical distributions in the 
$H \rightarrow Z Z^* \to 4\ell $ decay channel and in the VH and VBF 
production modes \cite{CP-review}. 

\begin{figure}[!h] 
\begin{center}
\vspace*{-.5mm}
\begin{tabular}{lll}
\begin{minipage}{5cm}
\hspace*{-1.6cm}
\epsfig{file=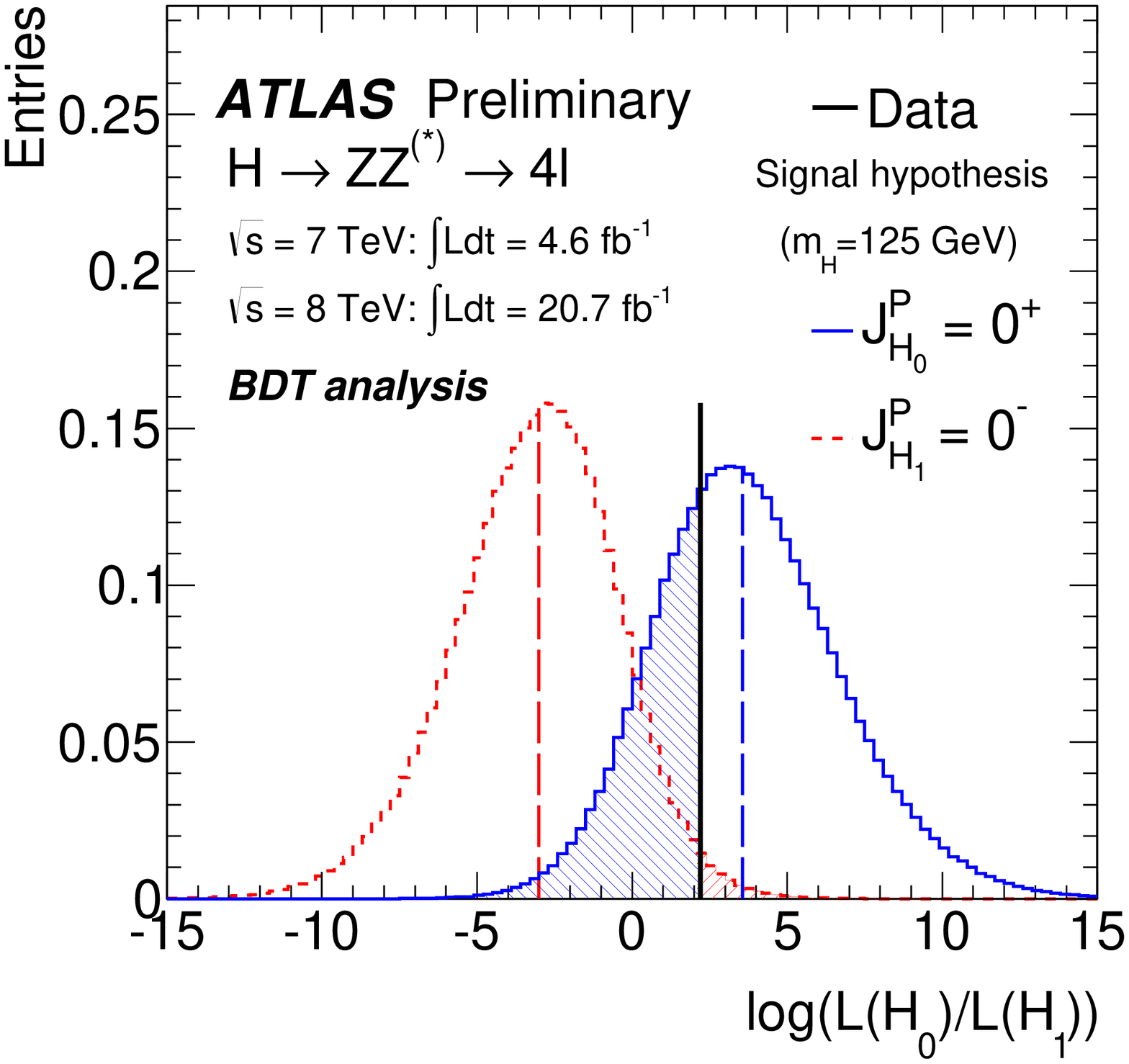,width=4.6cm}
\end{minipage}
& \hspace*{-2.2cm}
\begin{minipage}{5cm}
\epsfig{file=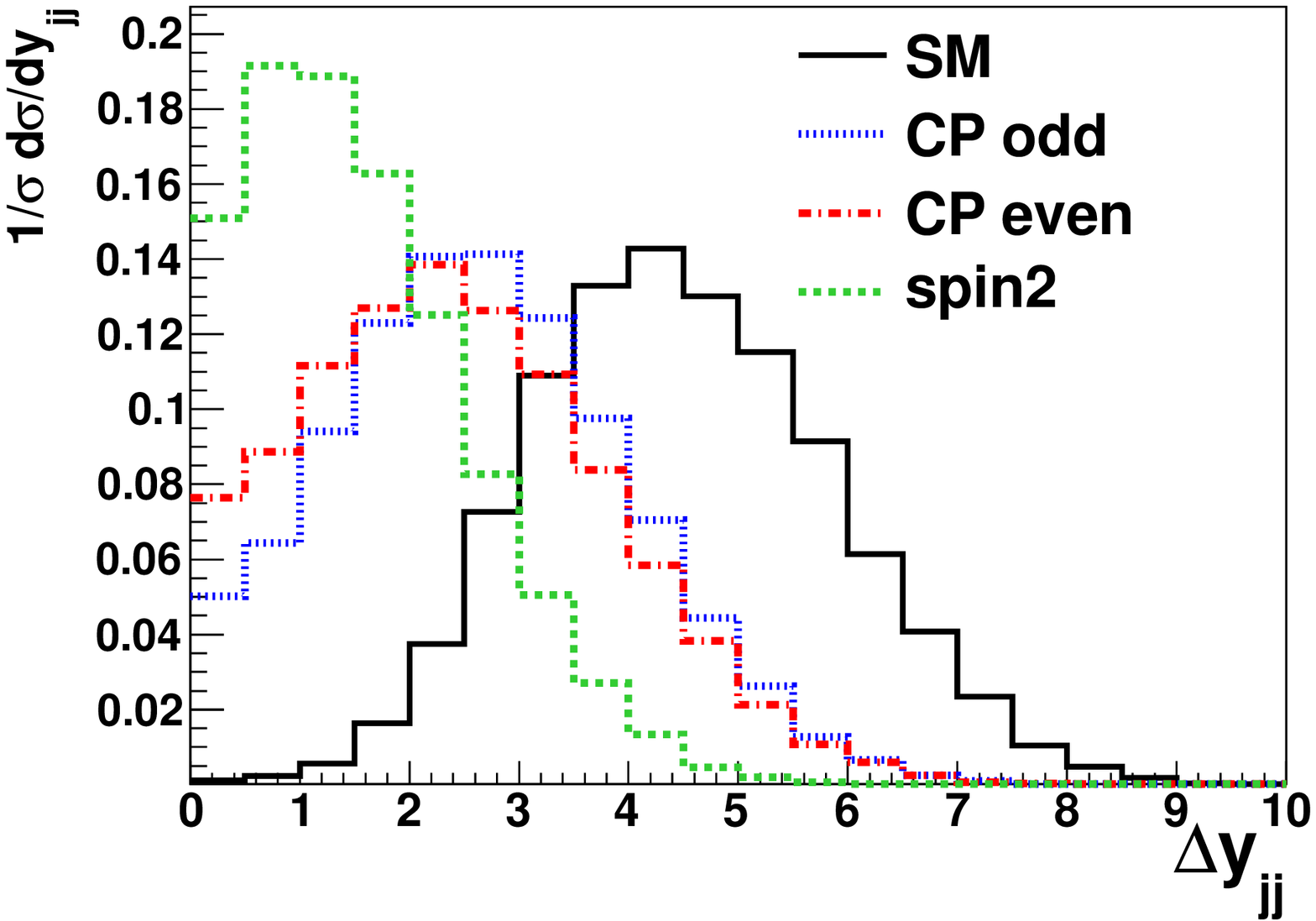,width=6cm}
\end{minipage}
& \hspace*{-2.2cm}
\begin{minipage}{5cm}
\epsfig{file=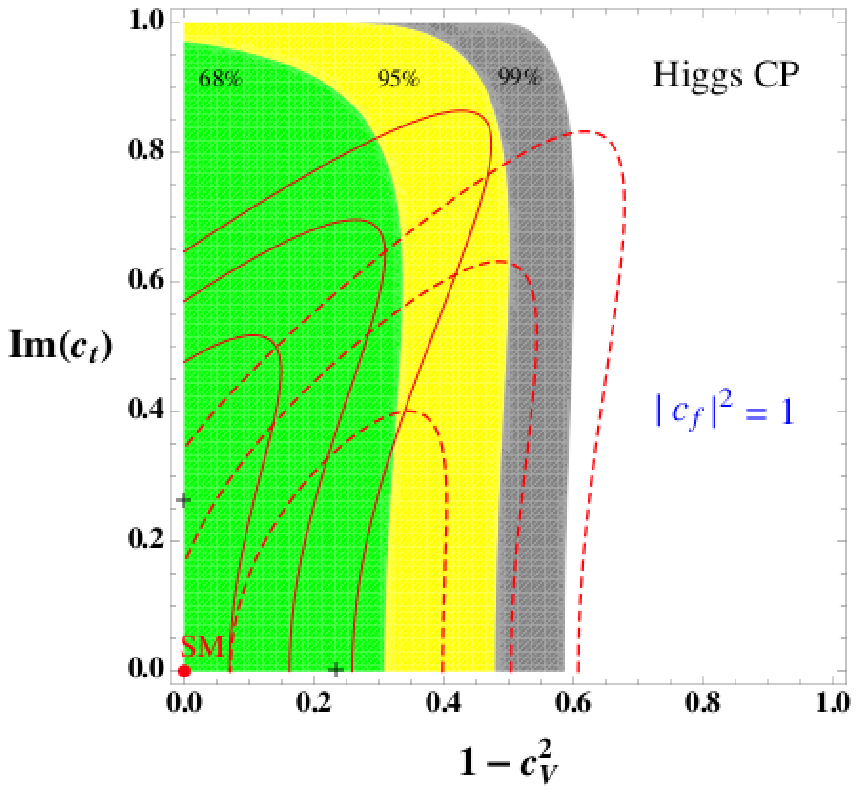,width=9.5cm}
\end{minipage}
\end{tabular}
\end{center}
\vspace*{-6mm}
\caption[]{
Left:discrimination between the $0^+$ and $0^-$ Higgs hypotheses
using the kinematics of the $H\to ZZ^* \to 4\ell$ channel by ATLAS \cite{CP-exp}.
Center: normalized distribution of the difference in rapidity between the 
scattered jets in VBF for  each of the SM and BSM operators (spin--2,  CP--even  
and CP--odd) individually \cite{CP-nous}.
Right: best-fit regions at $68\%$, $95\%$ and $99\%{\rm CL}$ 
in the plane $[1-c^2_V$,Im$(c_t)$ for $\vert c_t \vert^2\!=\!\vert c_f 
\vert^2\!=\!1$ \cite{fit}.
}
\label{fig:CP}
\vspace*{-2mm}
\end{figure}

With the 25 fb$^{-1}$ data collected so far, the ATLAS and CMS collaborations 
performed a matrix-element likelihood analysis which exploits the kinematics
and Lorenz structure of the $H\to ZZ^* \to 4\ell$ channel. it was found 
that the observed particle is more compatible with a $0^+$ state and the $0^-$ 
possibility is excluded at the 97.8\%CL or higher, as is shown in Fig.~\ref{fig:CP} 
(left) \cite{CP-exp}.  

Other useful diagnostics of the Higgs CP nature that
also rely on the tensorial structure of the $HVV$ coupling 
can be made in the VBF process. It was known since a long time that
in this channel, the distribution in the azimuthal angle between the two 
jets produced in  association with the Higgs discriminates a 
CP--even from a CP--odd state \cite{zepp}. This has been extended recently to other
observables, like the rapidity separation between the
two jets \cite{CP-nous}. An example of discrimination between CP-even and CP--odd 
Higgs couplings (and also spin 2) is shown in Fig.~\ref{fig:CP} (center). 

Nevertheless, there is a caveat in the analyses relying on the $HVV$ couplings: 
since a CP--odd state has no tree--level $VV$ couplings, the previous processes 
project out only the CP--even component of the $HVV$ coupling \cite{CP-tt} 
even if the state is a CP--even and odd mixture. Thus, in the CP studies above,  
one is simply verifying a posteriori that indeed the CP--even component is
projected out. In fact, a better way to measure the Higgs parity is to study the 
signal strength in the $H\to VV$ channels \cite{fit}. Indeed,  the $HVV$ coupling takes 
the  general form $g_{HVV}^{\mu \nu} = -i c_V (M_V^2/v)\; g^{\mu \nu}$
where $c_V$ measures the departure from the SM: $c_V\!=\!1$ for a pure $0^+$ 
state with SM--like couplings and $c_V\approx 0$ for a pure $0^-$ state.   The
measurement of $c_V$ should allow to determine the CP composition of the Higgs
and in all cases, the quantity  $1\!-\!c_V^2$ gives an {\it upper bound}
on the CP--odd contribution  to the $HVV$ coupling.
Using $\mu_{VV}$ 
and the ratios $\mu_{\gamma\gamma}/\mu_{VV}$ and $\mu_{\tau\tau}/\mu_{VV}$, it 
was demonstrated  that the observed Higgs has indeed a large CP component, 
$\gsim  50\%$ at the 95\%CL; see Fig.~\ref{fig:CP} (right) \cite{fit}. 


In fact, it turns out that the best and less unambiguous way to probe the CP nature of 
the Higgs boson would be to look at final states in which the particle decays 
hadronically. The best examples would be then to consider the processes 
$pp \to HZ \to b\bar b \ell \ell$ 
and $pp \to tt H \to tt bb$ and look for angular correlations \cite{CP-tt}. These 
processes are nevertheless extremely challenging.

\section{Implications for Supersymmetry}\vspace*{-1mm}   

We turn now to the implications of the LHC Higgs results for the MSSM Higgs sector, 
first from the mass value and the production rates and
then from heavy Higgs searches.  We start by discussing
the Higgs masses and couplings when
the important radiative corrections are included.\vspace*{-2mm}  

\subsection{The Higgs masses and couplings in the MSSM}\vspace*{-1mm}  

In the MSSM, the tree--level masses of the CP--even $h$ and $H$  bosons depend
only on $M_A$ and $\tb$. However, many parameters of the MSSM such as the 
masses of the third generation stop and sbottom squarks $m_{\tilde t_i},
m_{\tilde b_i}$  and their trilinear couplings $A_{t}, A_b$ enter $M_h$ and $M_H$ 
through quantum corrections.  These are introduced by a general $2\times 2$ matrix  
$\Delta {\cal M}_{ij}^2$ but the by far leading is controlled by the top Yukawa 
coupling which appears with the fourth power in the 22 entry \cite{CR-Higgs}
\beq
\label{higgscorr}
\Delta {\cal M}_{22}^2 \sim \! \frac{3 \bar{m}_t^4}{2\pi^2 v^2\sin^
2\beta} \left[ \log \frac{M_S^2}{\bar{m}_t^2} \!+ \! \frac{X_t^2}{M_S^2} \left( 1 \! -
\! \frac{X_t^2}{12M_S^2} \right) \right] \ , \ \ \lambda_t =  m_t/v \sin\beta
\eeq
where $M_S =\sqrt{ m_{\tilde{t}_1}m_{\tilde{t}_2}} $
is  defined to be the SUSY--breaking scale 
and $X_t\!= \! A_t\! - \! \mu/\tb$ the stop mixing parameter.
Other soft SUSY--breaking parameters enter but their contribution is small. 
The maximal value $M_h^{\rm max}$ is given in this approximation by 
$M_h^2 \stackrel{M_A \gg M_Z} \to  M_Z^2 \cos^2 2 \beta +
\Delta {\cal M}_{22}^2$ and is obtained for:
$i)$ a decoupling regime  $M_A\! \sim \mathcal{O}$(TeV);
$ii)$ large $\tb$ values;
$iii)$ heavy stops, i.e. large $M_S$; 
$iv)$ the so--called ma\-xi\-mal mixing scenario  $X_t=\sqrt{6}M_S$ \cite{benchmarks}.
If the parameters are optimized as above, the maximal $M_h$ value can reach the 
level of $130$ GeV.   

It was pointed out in Refs.~\cite{paper4,R1,Habemus} that when the measured value 
$M_h\!=\!125$ GeV is taken into account, the MSSM Higgs  sector with  only the largely 
dominant $\Delta {\cal M}_{22}^2$ correction, can be again described with 
only the two parameters $\tb$  and $M_A$; in other words,  the loop 
corrections are fixed by the value of $M_h$. This observation leads 
to a rather simple parametrisation of the MSSM Higgs sector, called  
$h$MSSM. One has 
for the heavier CP Higgs mass and the mixing angle $\alpha$
\begin{eqnarray}
\label{hMSSM-relation}
M_{H}^2 \! = \! \frac{(M_{A}^2+M_{Z}^2-M_{h}^2)(M_{Z}^2 c^{2}_{\beta}+M_{A}^2
s^{2}_{\beta}) - M_{A}^2 M_{Z}^2 c^{2}_{2\beta} } {M_{Z}^2 c^{2}_{\beta}+M_{A}^2
s^{2}_{\beta} - M_{h}^2} , \
\alpha\! =\! -\! \arctan\left(\frac{ (M_{Z}^2+M_{A}^2) c_{\beta} s_{\beta}} {M_{Z}^2
c^{2}_{\beta}+M_{A}^2 s^{2}_{\beta} - M_{h}^2}\right)~
\end{eqnarray}
while the charged Higgs mass is approximately given by its tree--level value $M_{H^\pm} \simeq \sqrt { 
M_A^2 + M_W^2}$. 
In Ref.~\cite{Habemus}, it has been shown that this $h$MSSM parametrisation is a very good approximation. 
 
In the MSSM, the couplings of the CP--even $h$ and $H$ bosons compared to the SM Higgs 
couplings depend on the angles $\beta$ and $\alpha$. In the case of $h$ outside the decoupling 
regime (where they reach unity), the reduced couplings to the heavy $t,b,\tau$  fermions 
and $V\! = \!W/Z$ bosons read
\begin{eqnarray}  
c_V^0  \! = \!  \sin(\beta \!-\! \alpha)  ,   \ c_t^0 \!  = \!  
\cos  \alpha/ \sin\beta ,   \ c_b^0 \!  = \!  -\!
\sin  \alpha/ \cos\beta  \label{Eq:MSSMlaws} 
\end{eqnarray}
If $\alpha$ is fixed by the $h$MSSM relation eq.~(\ref{hMSSM-relation}), then the $h$ 
couplings depend only on the inputs $\tan\beta, M_A$ as at tree--level. This is also the 
case of the heavier CP--even  $H$ boson. However, close to the decoupling regime that is indicated by the $h$ properties,  $H$  has approximately the same mass as the 
$A$ state and its interactions are similar. The MSSM Higgs spectrum will thus consist of 
a SM--like Higgs $h \equiv H_{\rm SM}$  and two pseudoscalar--like particles, $\Phi\! =\! 
H/A$. The $H^\pm$ boson will also be mass degenerate 
with the $\Phi$ states and the intensity of its couplings to fermions will be similar.  
In the high $\tb$ regime, the couplings of the heavy Higgs bosons to $b$ 
quarks and to $\tau$ leptons $\propto \tb$ are so strongly enhanced,  and the couplings  
to top quarks and $W/Z$ bosons $\propto 1/\tb$ suppressed, that the  pattern is rather  simple.
At very low $\tb$, the situation for the couplings is reversed.

\subsection{Implications from the Higgs mass value} 

We discuss now the implications of the measured mass value of the 
observed Higgs boson at the LHC, that we identify with the lightest state $h$ of
the MSSM, relying mostly on  Refs.~\cite{paper1,paper2}.  We will consider 
various options of the MSSM and Fig.~\ref{Fig:mass} below summarises the findings. 

\begin{figure*}[!h]
\vspace*{-4mm}
\begin{tabular}{lll}
\hspace*{-.8cm}
\begin{minipage}{5cm}
\vspace*{-3.cm}
\epsfig{file=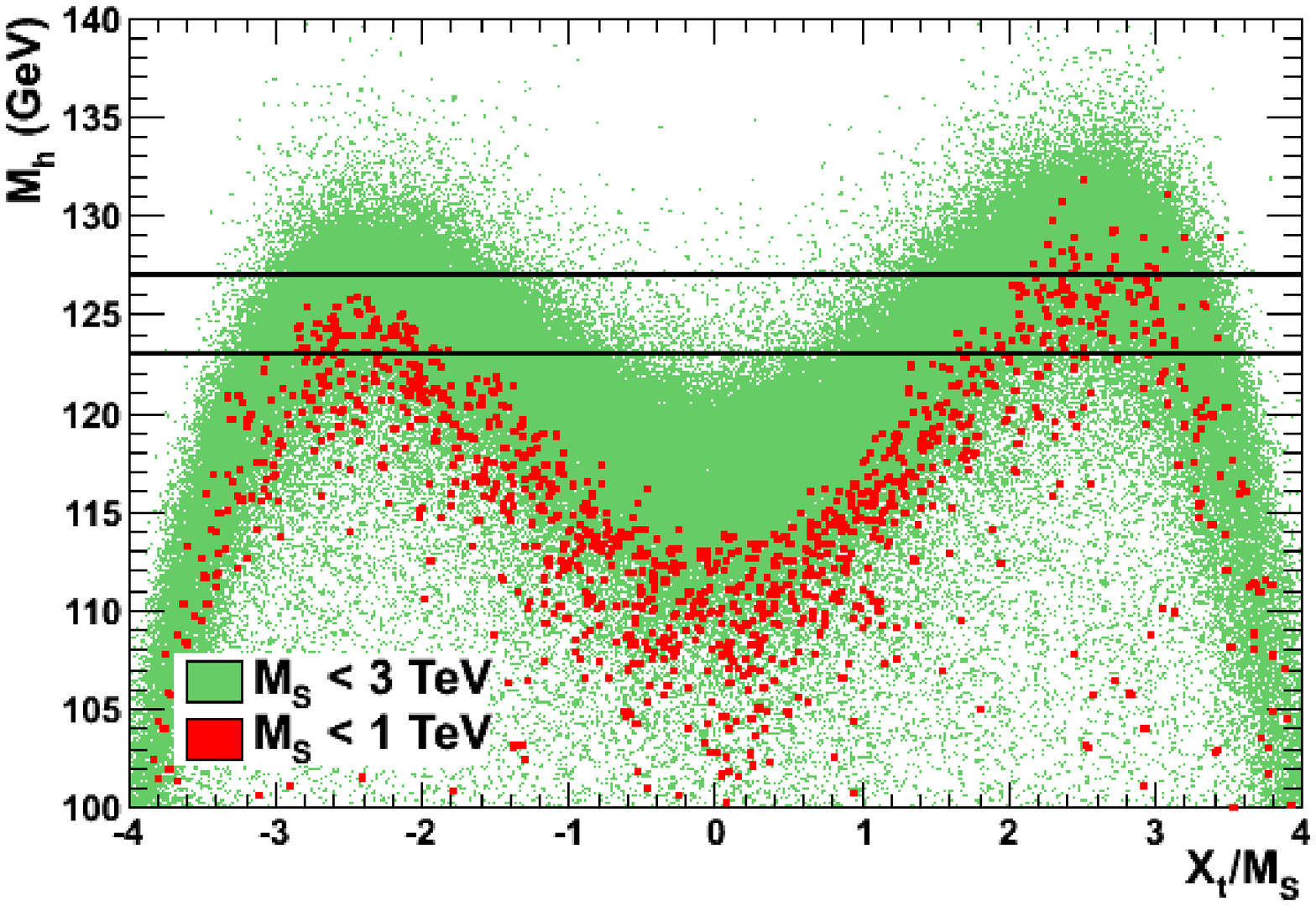, width=6.2cm}
\end{minipage}
& \hspace*{-.1cm}
\begin{minipage}{5cm}
\vspace*{-3.cm}
\epsfig{file=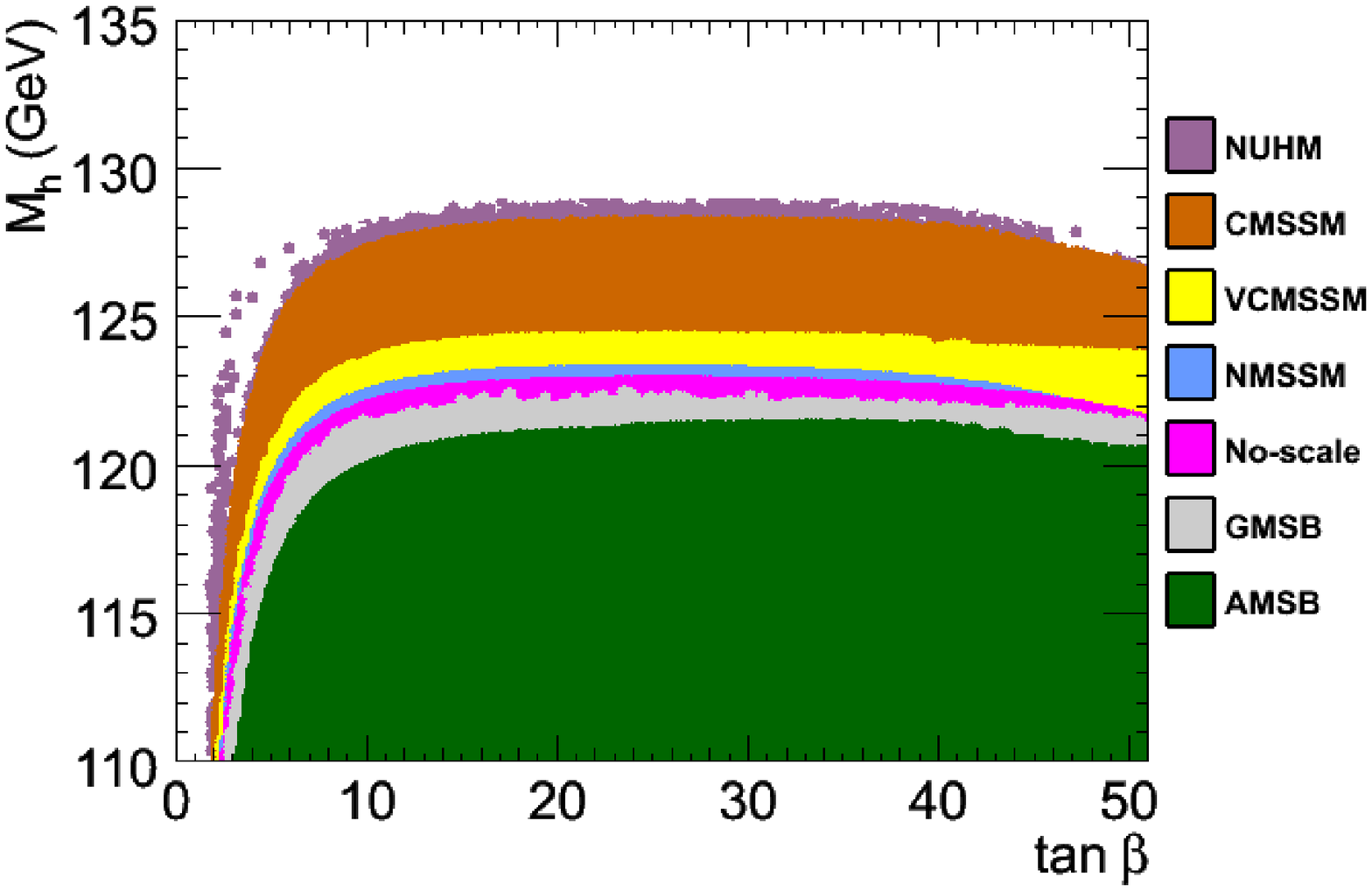, width=6cm}
\end{minipage}
& \hspace*{-.3cm}
\begin{minipage}{5cm}
\vspace*{.3cm}
\epsfig{file=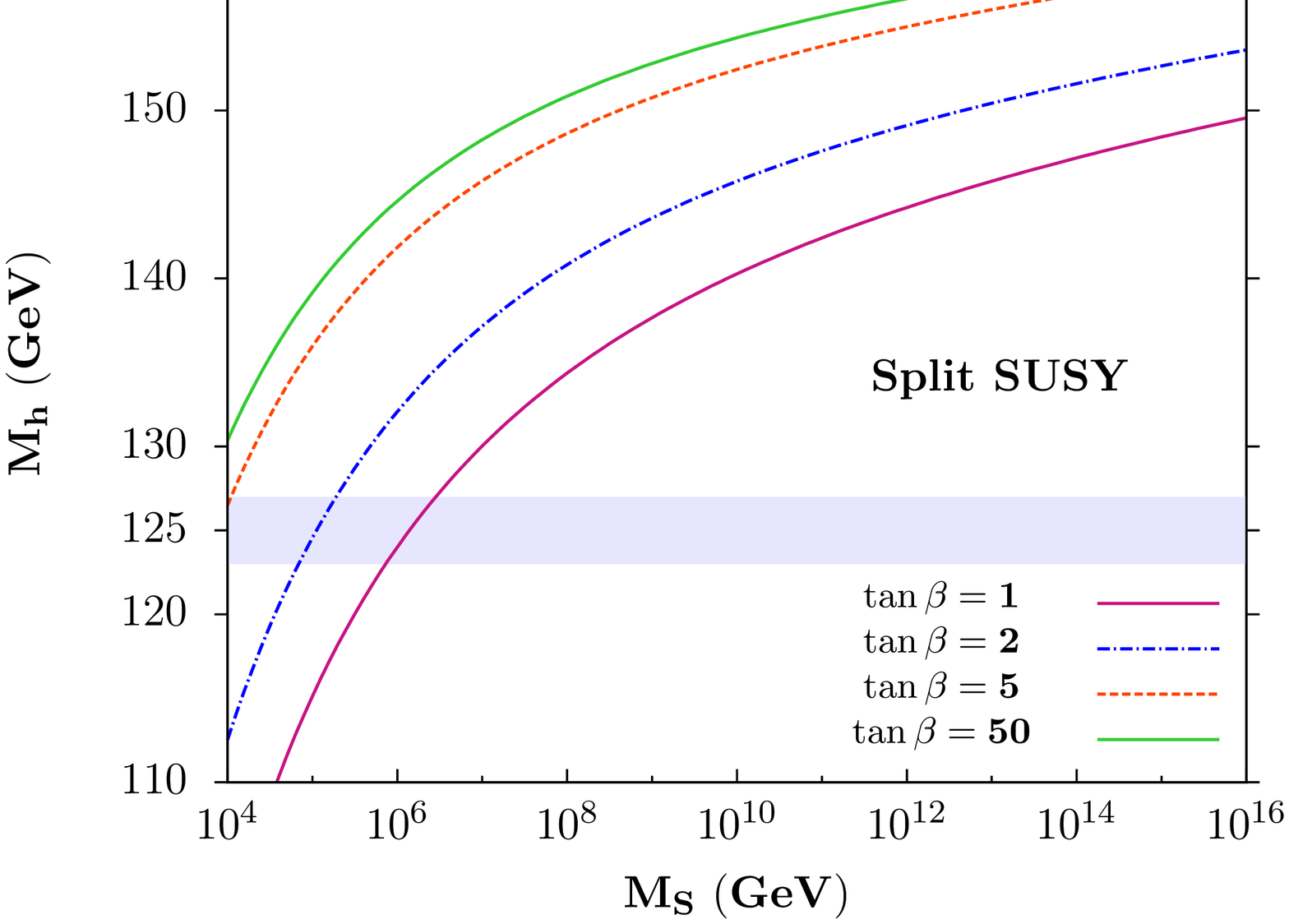, width=5cm}
\end{minipage}
\end{tabular} 
\vspace*{-3.5cm}
\caption{The maximal value of $M_h$ in the MSSM \cite{paper1}. Left: as a function of $X_t/M_S$ in the pMSSM 
when  all other SUSY parameters and $\tb$ are scanned. Center: as function 
of $\tb$ in various constrained MSSM scenarios. Right: as a function of $M_S$ in split SUSY
scenarios.}
\label{Fig:mass}
\vspace*{-1mm}
\end{figure*}

{\bf a) The phenomenological MSSM.} 
In  the so--called phenomenological MSSM (pMSSM)~\cite{pMSSM}, in which CP conservation,
flavour diagonal sfermion mass 
and coupling matrices and universality of the first and second sfermion generations 
are imposed, the model involves  only 22 free parameters in addition to those of the
SM: besides $\tb$ and $M_A$, these are the higgsino mass $\mu$,  the three gaugino 
masses $M_{1,2,3}$, the diagonal left-- and right--handed sfermion mass parameters 
$m_{ {\tilde f}_{L,R}}$ and the trilinear sfermion couplings $A_f$. One can then 
perform a scan of these parameters in a wide range to obtain the maximal
$M_h^{\rm max}$ value with the full radiative corrections included. 

Using the RGE program {\tt Suspect} \cite{Suspect} that calculates the particle spectrum 
in the MSSM, a large scan of the pMSSM parameters 
was performed \cite{paper1} and an illustration of  the results is shown in 
Fig.~\ref{Fig:mass} (left) where  the obtained $M_h^{\rm max}$ is displayed as a
function of the ratio $X_t/M_S$. The resulting values are
confronted to the mass range $123~{\rm GeV} \leq M_h \leq 127~{\rm GeV}$
when  the $\pm 2$ GeV parametric uncertainties from the
SM inputs and the  theoretical uncertainties in the determination of $M_h$ are included. 
For $M_S \! \lsim \! 1$ TeV, only the scenarios with $X_t/M_S$ values close to maximal  
mixing $X_t/M_S \approx \sqrt 6$ survive. The no--mixing scenario $X_t \approx 0$ 
is ruled out for $M_S \lsim 3$ TeV, while the typical mixing scenario, $X_t \approx 
M_S$, needs large $M_S$ and moderate to large $\tan\beta$ values.  

Note that masses above 1 TeV for squarks and gluinos are also required by the direct 
searches of SUSY particles at the LHC \cite{Craig} confirming the need of high $M_S$ values. Nevertheless, relatively light stops 
are still possible: as $M_S\! = \! \sqrt { m_{\tilde t_1} m_{\tilde t_2}}$, the possibility that 
$m_{\tilde t_1}$ is of the order of a few 100
GeV is still allowed, provided that stop mixing (leading to a significant
$m_{\tilde t_1}, m_{\tilde t_2}$ splitting) is large. Light 
electroweak sparticles such as sleptons, charginos and neutralinos 
are also still possible allowing for a ``natural SUSY" scenario \cite{natural} despite of
the large value  $M_h \approx 125$ GeV. 
 
{\bf b) Constrained MSSM scenarios.}
In constrained scenarios (cMSSM),  the various soft SUSY--breaking parameters 
are a result of the RGE evolution and obey universal boundary conditions at a high scale, 
thus reducing the number of basic input parameters to a handful. Three classes of such models 
have been widely discussed \cite{SUSY}: the minimal supergravity (mSUGRA) model  in 
which SUSY--breaking occurs in a hidden sector which communicates with the visible sector 
only via flavour-blind gravitational interactions leading to  common $m_{1/2}, m_0, A_0$
values for the gaugino masses, sfermion masses and sfermion trilinear couplings. 
Then come the gauge mediated and anomaly mediated SUSY--breaking scenarios, GMSB and AMSB,   
in which SUSY--breaking is communicated to the visible sector via gauge interactions and a  
super-Weyl anomaly. These models are described by $\tb$, the sign of $\mu$ and a few continuous parameters.

Again, a large scan of these input parameters has been performed in Ref.~\cite{paper1,paper2}
with the constraint that  $M_S \lsim 3$ TeV not to allow for too much fine-tuning 
\cite{natural}, and the results are shown in Fig.~\ref{Fig:mass} (center) for several cMSSMs  as a function of $\tan\beta$, the input that is common to all models. The resulting 
$M_h^{\rm max}$ is 121 GeV in AMSB and 122 GeV in GMSB, which means that these two scenarios are disfavoured if $M_S \lsim 3$ TeV. In mSUGRA, one obtains $M_h^{\rm max} \! = \! 128$ GeV and, thus, some parameter space would still survive; but in more constrained versions, the limit is lower. These limits can be qualitatively understood by considering the allowed values of the stop coupling $A_t$ which in many cases cannot be large compared to $M_S$ and hence does not allow maximal mixing. 

{\bf c) Split and high--scale SUSY models.} 
The choice $M_S \lsim 3$ TeV made in the previous discussion is mainly
dictated by fine--tuning considerations which are rather subjective. 
One could well abandon the SUSY solution to the hierarchy problem and 
have a very high $M_S$ implying  that, except for the $h$ boson, all 
other scalar particles are inaccessible. This is the split SUSY scenario 
in which the soft SUSY--breaking scalar mass terms except for one Higgs 
doublet are  assumed to have a common value $M_S \gg 1$ TeV. If the 
mass parameters for the spin--$\frac12$ gauginos and higgsinos are also 
assumed to be large, we are then  in a high scale SUSY scenario. See 
Ref.~\cite{high-scale} for reviews. 

In the Higgs sector, the main feature of these scenarios is that at the 
scale $M_S$,  the boundary condition on the quartic Higgs coupling  
is determined by SUSY, $\lambda(M_S)\! =\! \frac{1}{4}[ g^2(M_S)\!+\!
g^{\prime 2} (M_S)]\cos^22\beta$. But if the scalars are heavy, they lead to 
radiative corrections that are significantly  enhanced by large $\log( 
M_S/M_Z)$ terms. In order to have 
reliable predictions, one needs to properly decouple the heavy states from the 
low-energy theory and resum the large logarithms. This work has been done in e.g. 
Ref.~\cite{bds} and the results are implemented in the  code {\tt SuSpect} \cite{Suspect}. 

Using this tool, a scan in the $[\tan\beta, M_S]$ plane  has been performed to 
determine the value of $M_h$ in the split and high--scale SUSY scenarios. 
The requirement that 123 $\lsim M_h \lsim$ 127 GeV imposes strong constraints on the 
parameters of these two models. For this mass range and $\tan\beta\approx 1$, very 
large scales are needed in the high--scale SUSY scenario, while scales $M_S\! 
\approx \! 10^{4}~{\rm GeV}$ are required in the split--SUSY case; see 
Fig.~\ref{Fig:mass} (right). In the latter case, to cope with $M_h =125$ GeV, 
SUSY should manifest itself at much smaller scales at $\tan\beta\! \gg \! 1$, $M_S \approx 1$ TeV  as seen previously.  

{\bf d) Splitting the Higgs and sfermion sectors.}
In these high SUSY scale scenarios, the Higgs mass parameters were assumed 
to be related to the mass scale of the scalar fermions in such a way that
the masses of the heavier Higgs particles are also large, $M_A 
\approx M_S$. However,  this needs  not to be true in general and one 
may be rather conservative and allow any value for $M_A$ 
irrespective of $M_S$. This is the quite ``model--independent" approach that 
has been advocated in Refs.~\cite{paper4,R1,Habemus}: 
take $M_A$ as a free parameter of the pMSSM, with values ranging from ${\cal O}( 
100$ GeV) up to ${\cal O}(M_S)$. 

An important consequence of this possibility is that it reopens the low
$\tan\beta$ region, $\tan\beta \! \lsim \! 3$, that was long thought to be forbidden
if one requires a SUSY scale $M_S \! \lsim \! 1$ TeV \cite{benchmarks}, as a result of the LEP2 limit 
$M_h \! \gsim \! 114$ GeV \cite{PDG}. Values $\tb \approx 1 $ with a correct $M_h$ 
are possible  provided $M_S$ is large enough, e.g. $M_S \! \approx \! 20$ TeV for
$\tb \! \approx \! 2$. In this scenario, one can use the $h$MSSM discussed in the 
previous subsection in which the Higgs sector with $M_h=125$ GeV is again described by $M_A$ and $\tb$
at higher orders, which makes the discussion rather simple and transparent.

\subsection{Implications from the production rates of the observed Higgs state}

In the MSSM, the couplings of the Higgs particles depend in principle 
only on the two  inputs $[\tan\beta, M_A]$ as discussed previously. 
However, this is true when only the radiative corrections to the Higgs masses 
and to the angle $\alpha$ are included. There are also direct corrections to the couplings  
which  alter  this simple picture and there are two important ones. 
First, for $b$--quarks, additional one--loop vertex corrections modify 
the tree--level Higgs--$b \bar b$ coupling: they grow as $ m_b \mu \tan\beta$
and can be very large at high  $\tb$. These are the famous $\Delta_b$ corrections
\cite{deltab} which,  outside the decoupling regime, 
make that the coupling $g_{h b \bar b} \propto c_b$ reads
\beq
c_b \approx c_b^0 \times [1- \Delta_b/(1+\Delta_b) \times (1+ \cot\alpha \cot\beta)]  
\label{cb}\!
\eeq
with $\tan\alpha\! \to\! -\! 1/\tb$ for $M_A\! \gg\! M_Z$. A large $\Delta_b$  would 
significantly alter the dominant  $h\! \to b\bar b$ partial width and affect 
the  branching fractions of  all other decays. 

\begin{figure}[!h]
\begin{center}
\vspace*{-5.3cm}
\begin{tabular}{ll}
\hspace*{-.8cm}
\begin{minipage}{8cm}
\epsfig{file=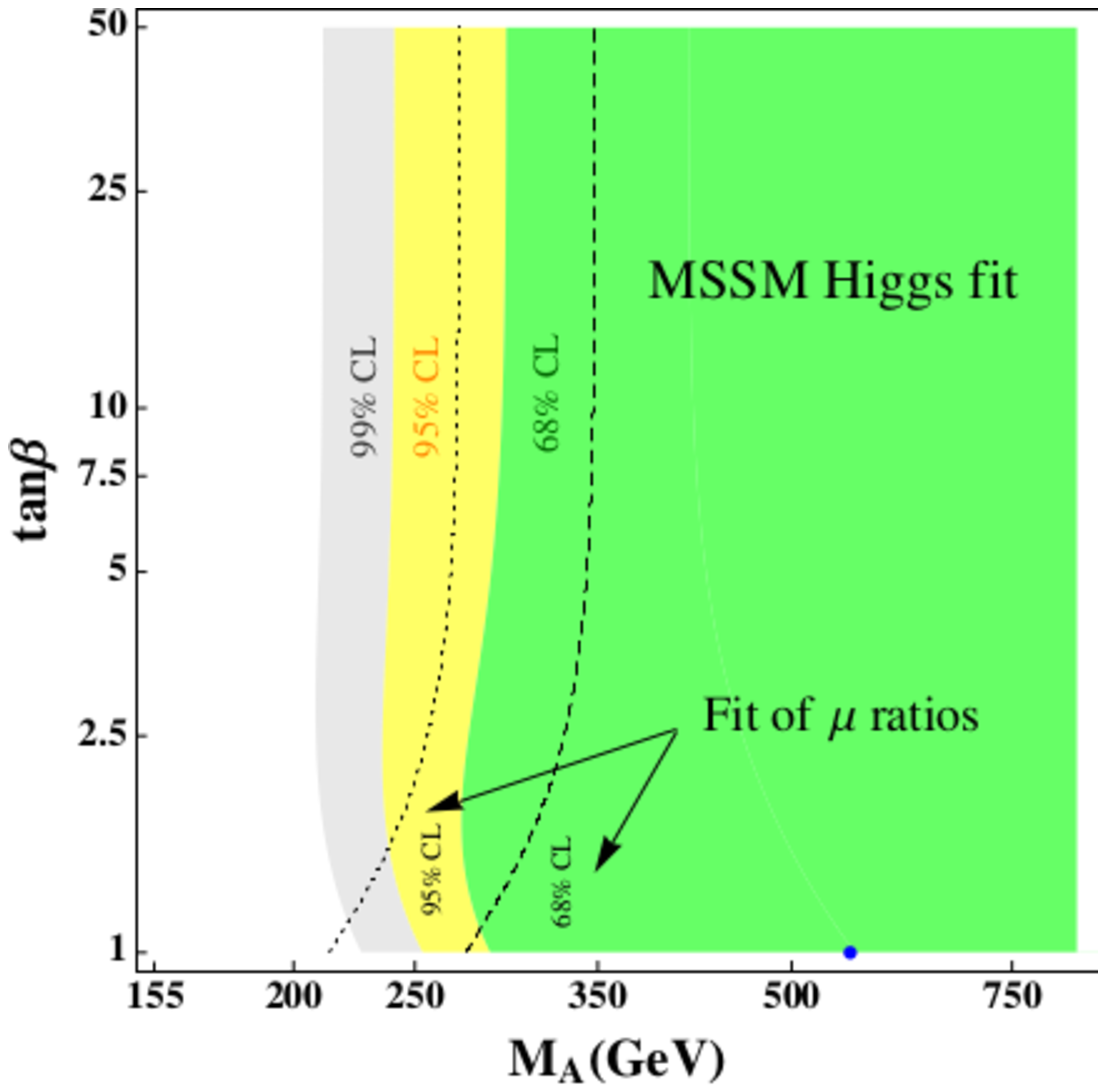,width=9cm}
\end{minipage}
& \hspace*{-2.5cm}
\begin{minipage}{8cm}
\epsfig{file=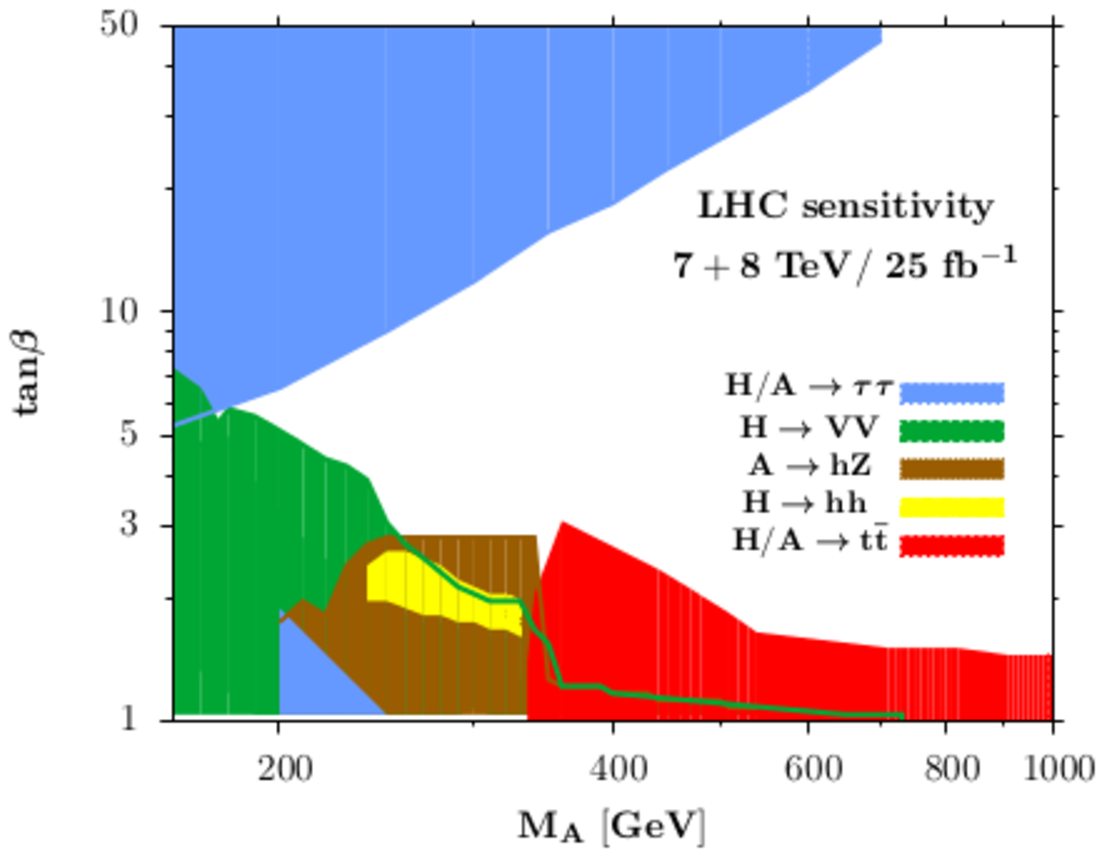,width=11cm}
\end{minipage}
\end{tabular}
\vspace*{-5.7cm}
\caption{Left: best-fit regions for the signal strengths and their ratios in the 
plane $[\tan \beta, M_A]$; the best point is in  blue \cite{Habemus}. Right: the estimated sensitivities 
in the various search channels for the heavier MSSM Higgs bosons in the $[\tb,M_A]$ plane
with the 25 fb$^{-1}$ data collected so far at the 7+8 TeV LHC \cite{paper4}.}
\label{Fig:sensitivity}
\vspace*{-5mm}
\label{tbMA}
\end{center}
\end{figure}

Second, the $h t\bar t$ coupling is derived indirectly from the $gg\! \to\! h$ 
production cross section and the $h\! \to\! \gamma \gamma$ decay branching ratio,
two processes that are generated by triangular loops. In the MSSM, these loops 
involve not only the top quark (and the $W$ boson in the decay $h\to \gamma \gamma$) 
but also contributions from not too heavy superparticles. In particular, there are stop
loops  that alter significantly the two process and change the coupling $c_t^0$ to
\cite{Stop}
\beq 
c_t \! \approx \! c_t^0 [ 1 \!+\! m_t^2/( 4 m_{\tilde t_1}^2 
m_{\tilde t_2}^2 ) \times  ( m_{\tilde t_1}^2 \! + \! m_{\tilde t_2}^2  \! - \! 
X_t^2) ] 
\label{ct}
\eeq
Hence, because of these corrections,  the Higgs should be characterized at the LHC 
by at least the three independent  $h$ couplings $c_t$, $c_b$ and $c_V=c_V^0$ 
as discussed in section 2.2. The previously discussed fit, which allows to derive
best-fit values $c_t=0.89, ~ c_b=1.01$ and $c_V=1.02$ of Fig.~\ref{tbMA} can be
used to constrain these corrections. In turn,   ignoring the direct corrections
(which are small for a high enough $M_S$) and  using the 
input $M_h \! \approx 125$ GeV, one can make a fit in the plane $[\tan
\beta,  M_A]$. This is shown in the left--hand side of Fig.~\ref{tbMA} where
the 68\%, 95\% and  99\%CL contours from the signal strengths and their ratios
are displayed when  the theory uncertainty is taken as a bias. 
The best-fit point is obtained  $\tan\beta\!=\! 1$  and $M_A \! = \! 550 
\; {\rm GeV}$ (which implies a large SUSY scale, $M_S = {\cal O}(100)$ TeV).
Larger values of $\tb$ and lower values of $M_A$ could also
be accommodated reasonably well by the fit, allowing thus for not too
large $M_S$. In all, cases one has $M_A \gsim 200$--350 GeV though.

\subsection{Implications from heavy Higgs boson searches}\vspace*{-2mm}

We turn now to the constraints on the MSSM Higgs sector that can be obtained
from the search of the heavier $H/A$ and $H^\pm$ states at the LHC. At high 
$\tb$ values, the strong enhancement of the $b,\tau$  couplings makes that 
the  $\Phi=H/A$ states  decay dominantly into  $\tau^+\tau^-$ 
and  $b\bar b$ pairs,  with branching ratios of BR$(\Phi \to \tau \tau) \approx  
10\%$ and BR$(\Phi \to b \bar b) \approx 90\%$, while the decays $\Phi \to t\bar t$
are negligible. The $H^\pm$ boson decays into $\tau \nu_{\tau}$ final 
states with a branching fraction of almost 100\% for $H^\pm$ masses  below the $tb$
threshold, $M_{H^\pm} \lsim m_t+m_b$, and a branching ratio of only $\approx
10\%$ for masses above this threshold while the rate for $H^\pm  \to t b$ will
be at the $\approx 90\%$ level in most cases.  Concerning the production,  only two 
processes are relevant in this case:  $gg\!  \to\!  \Phi$ fusion with the $b$--loop 
included \cite{ggH-NLO} and associated production with $b$--quarks, $gg/q\bar q \! 
\to \! b\bar b \! +\! \Phi$ \cite{bbH-NNLO}. The most powerful LHC search channel 
is thus $pp\! \to \!gg\! +\! b \bar b \! \to \! \Phi \! \to \! \tau^+ \tau^-$. 
For the charged Higgs, the dominant mode is $H^\pm \! \to \!  
\tau \nu$  with the $H^\pm$ light enough to be produced in top decays
$t \! \to \! H^+ b \! \to \! \tau\nu  b$. 

In the low $\tb$ regime, $\tb \lsim 3$, the phenomenology of the $A,H,H^\pm$ states  
is richer \cite{paper4}. Starting with the production,  there is no Higgs strahlung
and VBF processes in the decoupling limit while the rates for associated $t\bar t \Phi$ 
and $b\bar b \Phi $ are small because the $\Phi tt\;(\Phi bb)$ couplings are suppressed (not 
sufficiently enhanced).  Only the $gg\to \Phi $ process with the dominant $t$ and 
sub-dominant $b$ contributions included  provides large rates. For $H^\pm$, the 
dominant production channel is again top quark decays, $t \to H^+ b$ for $M_{H^\pm} 
\lsim 170$ GeV; for higher $H^\pm$ masses,  the main process is $gg/q\bar q  \to H^\pm tb$. 
Turning to the $H/A/H^\pm$ decay pattern,  it can be rather involved  for $\tb \lsim 3$.
Above the $t\bar t\; (tb)$ threshold for $H/A(H^\pm)$,  the channels 
$H/ A \rightarrow t\bar{t}$  ($H^+ \to t \bar b$) are by far dominant for 
and do not leave space for any other mode. 
Below the $t\bar t$ threshold, the  $H\! \to \! WW,ZZ$ decay rates are still 
significant as $g_{HVV}$ is not completely suppressed.  
For $2M_h \lsim M_H \lsim 2m_t$, $H\to hh$ is the dominant $H$ decay mode 
as the $Hhh$ self--coupling is large at low $\tb$.
For $M_A\! \gsim M_h+M_Z$, $A \to hZ$ decays would occur but the $A \to \tau\tau$ 
channel   is still important with rates  $\gsim\! 5\%$. 
In the case of $H^\pm$, the channel  $H^+\! \to \! Wh$ is important for 
$M_{H^\pm}\! \lsim \! 250$ GeV, similarly to the $A \! \to \! hZ$ case. 

In Ref.~\cite{paper4} a preliminary analysis of these channels has been 
performed using current information given by the ATLAS and CMS collaborations
in the context of searches for the SM Higgs boson or other heavy resonances. 
The results are shown in Fig.~\ref{Fig:sensitivity} (right) with an extrapolation 
to the full 25 fb$^{-1}$ data of the 7+8 TeV LHC run (it has been assumed  that the
sensitivity scales simply as the  square root of the number of events). The
sensitivities from the usual $H/A \to \tau^+\tau^-$  and $t \to bH^+ \to b \tau
\nu$ channels directly taken from the ATLAS and CMS analyses \cite{H-LHC} are also shown. 
The green and red areas correspond to the domains
where the $H\to VV$ and $H/A \to t\bar t$ channels become   constraining. The 
sensitivities in the $H\to hh$ and $A\to hZ$ modes are  given by
the yellow and brown areas which peak in the mass range $M_A=250$--350 GeV 
at low $\tb$ values. 

The outcome of the searches is impressive. The ATLAS and CMS $H/A \to \tau^+ \tau^-$ constraint is extremely restrictive and for $M_A \lsim 250$ GeV, it excludes  almost the entire
intermediate and high  $\tb$ regimes.   The constraint is less effective for a heavier 
$A$ but even for $M_A \approx 400$ GeV the high $\tb \gsim 10$ region is excluded and 
one is even sensitive to large values  $M_A \approx 800$ GeV for  $\tb \gsim 50$. 
For the charged Higgs boson, almost the entire $M_{H^\pm} \lsim 160$ GeV region 
is excluded by the process $t \to H^+ b$ with the decay $H^+ \to \tau \nu$.
The other channels, in particular $H \to VV$ and $H/A
\to t \bar t$, are very constraining as they cover the entire low    
$\tb$ area that was previously excluded by the LEP2 bound up to $M_A \approx
500$ GeV. Even  $A \to hZ$ and $H \to hh$ would be visible at the current LHC 
in small portions of the low $\tb$ parameter space.

\section{Perspectives for Higgs and new physics}\vspace*{-2mm}

The last few years were extremely rich and exciting for particle physics. With the 
historical discovery of a Higgs boson by the LHC collaborations ATLAS and CMS,   
crowned by a Nobel prize in fall 2013, and the first probe of its basic properties,
they witnessed a giant step in the unraveling of the mechanism that breaks the 
electroweak symmetry and generates the fundamental particle masses. They promoted 
the SM as the appropriate theory, up to at least the Fermi energy scale, to describe 
three of Nature's interactions, the electromagnetic, weak and strong forces. 
However, it is clear that these few years have also led to some frustration as no 
signal of physics beyond the SM has emerged from the LHC data. The hope of observing 
some signs of the new physics models that were put forward to address the 
hierarchy problem, that is deeply rooted in the Higgs mechanism, with  Supersymmetric 
theories being the most attractive ones, did not materialize. 

The Higgs discovery and the non--observa\-tion of
new particles has nevertheless far reaching consequences for supersymmetric theories 
and, in particular, for their simplest low energy formulation, the MSSM. The mass of 
approximately 125 GeV of the observed Higgs boson implies that the scale of 
SUSY--breaking is rather high, at least ${\cal O}$(TeV). This is backed up by
the limits on the masses of strongly interacting SUSY particles set by  
the ATLAS and CMS searches, which in most cases exceed the TeV range.  This
implies that if SUSY is indeed behind the stabilization of the Higgs mass against 
very high scales that enter via quantum corrections, it is either
fine--tuned at the permille level at least or its low energy manifestation
is more complicated than expected. 
 
The production and decay rates of the observed Higgs particles, as well as its
spin and parity quantum numbers, as measured by ATLAS and CMS 
with the $\approx 25$ fb$^{-1}$ data collected at $\sqrt s\!=\!7$+8 TeV, 
indicate that its couplings to fermions and gauge bosons are almost SM--like. 
In the context of the MSSM, this implies that we are close to the decoupling regime 
and this particle is the lightest $h$ boson, 
while the other $H/A/H^\pm$ states must be heavier than approximately the Fermi 
scale. This last feature is also backed up by LHC direct searches of 
these heavier Higgs states. 

This drives up to the question that is now very often asked: what to do next? The answer is, for me,  obvious: we are only in 
the beginning of a new era. Indeed, it was expected 
since a long time that the probing of the electroweak symmetry breaking mechanism 
will be at least a two chapters story. The first one is the search and the 
observation of a Higgs--like particle that will confirm the scenario of the SM and 
most of its extensions, that is, a spontaneous symmetry breaking by a scalar field 
that develops a non--zero vev. This long chapter has just 
been closed by the ATLAS and CMS collaborations with the spectacular observation 
of a Higgs boson. This observation opens a second and equally important chapter: 
the precise determination of the Higgs profile and the unraveling of the electroweak 
symmetry breaking mechanism itself. 

A more accurate measurement of the Higgs couplings to fermions and gauge bosons 
will be mandatory to establish the exact nature of the mechanism 
and, eventually, to pin down effects of new physics if additional ingredients beyond 
those of the SM are involved. This is particularly true in weakly interacting theories
 such as SUSY in which the quantum effects are expected to be small.
These measurements could be performed at the upgraded LHC with an energy close 
to $\sqrt s\!=\!14$ TeV, in particular if a very high luminosity, a few ab$^{-1}$, 
is achieved \cite{H-LHC,Snowmass}. 

At this upgrade, besides improving the measurements performed so far, rare but important 
channels such as associated Higgs production with top quarks, $pp\!\to\! t\bar t H$, 
and Higgs decays into $\mu^+ \mu^-$ and $Z\gamma$ states could be probed. Above all,  
a determination of the self--Higgs coupling could be made by searching for double Higgs 
production e.g. in the gluon fusion channel $gg\to HH$ \cite{HHH}; this would
be a first step towards the reconstruction of the scalar potential that is responsible 
of electroweak symmetry breaking.  This measurement would be difficult at the LHC 
even with high--luminosity but
a proton collider with an energy $\sqrt s\!=\!30$ to 100 TeV could do the job \cite{Snowmass}.

\begin{figure}[!h]
\vspace*{-2cm} 
\begin{tabular}{lll}
\hspace*{-1.4cm} 
\begin{minipage}{8cm}
\epsfig{file=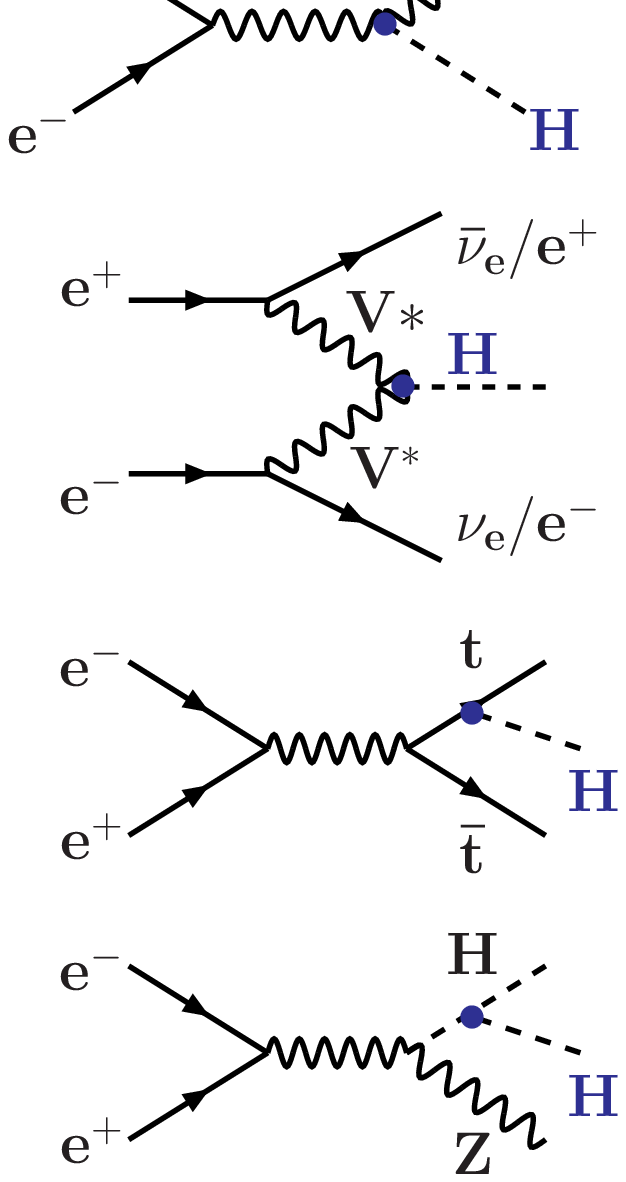,width=6cm}
\end{minipage}
& \hspace*{-6cm}
\begin{minipage}{8cm}
\vspace*{-.1cm}
\epsfig{file=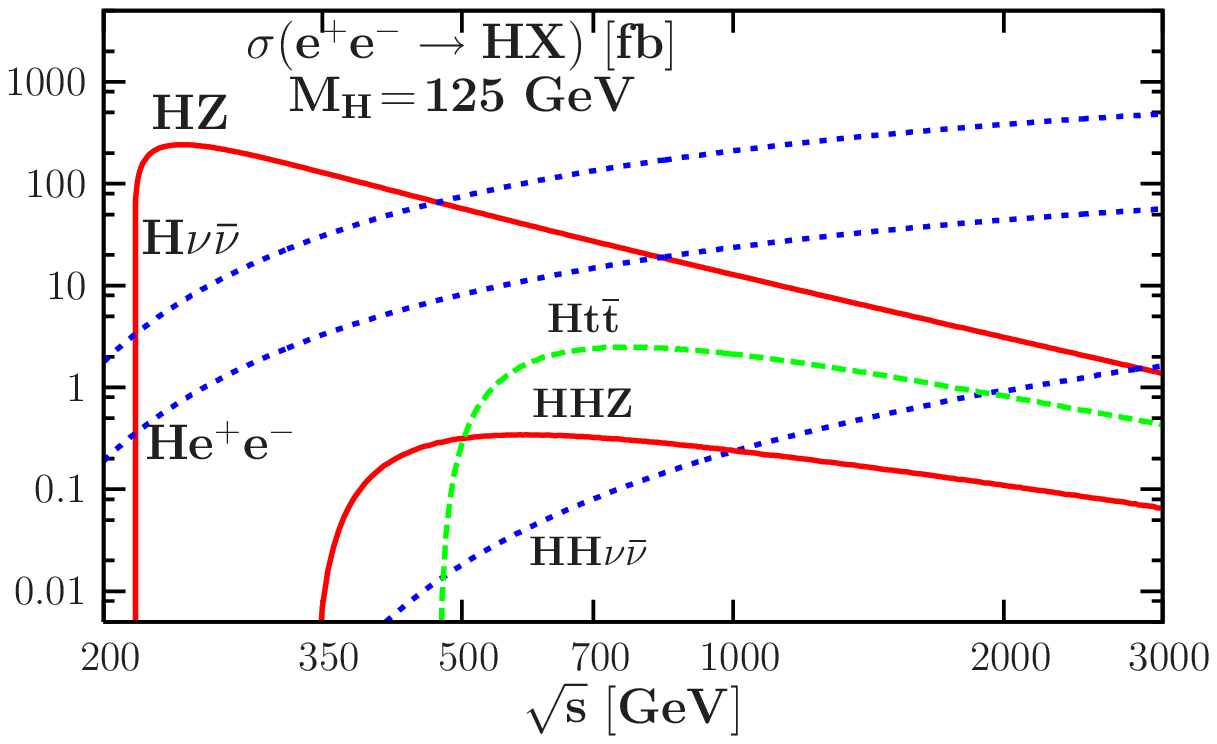,width=10cm,height=18cm}
\vspace*{-5cm}
\end{minipage}
& \hspace*{.4cm}
\begin{minipage}{6cm}
\vspace*{-4.5cm}
\renewcommand{\arraystretch}{.7}
\hspace*{2mm}
\begin{tabular}{|lc|} \hline
$ \Delta g_{HWW}$ & $ \approx \pm 0.01 $ \\
$ \Delta  g_{HZZ}$&  $\approx \pm 0.01 $  \\
$ \Delta  g_{Hbb}$& $ \approx \pm 0.02 $ \\
$ \Delta  g_{Hcc}$& $ \approx \pm 0.04 $ \\
$ \Delta  g_{H\tau\tau}$ & $\approx \pm 0.03 $  \\
$ \Delta g_{Htt}$& $\approx \pm 0.03 $  \\
$ \Delta  \lambda_{HHH}$& $ \approx \pm 0.20  $ \\ 
$ \Delta  M_H$ & $\approx \pm$ 40 MeV  \\
$\Delta  \Gamma_H^{\rm tot}$  & $\approx \pm 0.06$  \\
$\Delta {\rm CP}$ & $\approx \pm 0.04$ \\ \hline
\end{tabular}
\vspace*{-2mm}
\vspace*{1.5cm}
\end{minipage}
\end{tabular}
\vspace*{-7.3cm}
\caption{Diagrams for Higgs production in $e^+e^-$ collisions (left),  production
cross sections for a 125 GeV SM--Higgs as a function of $\sqrt s$ (center) and 
expected accuracies (right) for various coupling measurements (as well as mass, total 
width and CP--odd mixture) at the ILC with energy up to $\sqrt s=500$ GeV and 
a luminosity of up to 500 fb$^{-1}$ (see the discussion in G. Aarons et al. in Ref.~
\cite{ILC}).}
\label{Fig:ILC} 
\vspace*{-2mm}
\end{figure}

In a less near future, a high--energy lepton collider, which is nowadays discussed in various 
options (ILC, TLEP, CLIC, $\mu$--collider)  would lead to a more accurate probing of the Higgs properties \cite{ILC}, promoting the scalar sector to  the very high--precision level of the 
gauge and fermion sectors achieved by the LEP and SLC colliders in the 1990s \cite{PDG}. 
In $\protect{e^{+} e^{-}}$ collisions, the process 
$ee \to HZ$, just looking at the recoiling $Z$ boson
allows to measure the Higgs mass, the CP parity and the absolute $HZZ$ coupling,  allowing 
to derive the total decay width $\Gamma_H^{\rm tot}$. One can then measure precisely, 
already at $\sqrt s \approx 250$ GeV where $\sigma(e^+ e^- \to HZ)$ is maximal, the 
absolute Higgs couplings to gauge bosons and light fermions at the \% level from the 
decay branching ratios. The important couplings to top quarks and the Higgs self--couplings 
can measured at the 10\% level in the higher-order processes $e^+ e^- \to t\bar t H$ and 
$e^+ e^- \to HHZ$ at energies of at least 500 GeV with a high--luminosity. Some information 
on this issue is given in Fig.~\ref{Fig:ILC}.

Besides the high precision study of the already obser\-ved Higgs, one
should also continue to search for the heavy states that are predicted by
SUSY, not only the superparticles but also the heavier Higgs bosons. 
The energy upgrade to  $\approx \!14$ TeV (and eventually beyond) and the planed 
order of magnitude (or more) increase in luminosity will allow to probe much higher
mass scales than presently.  In fact, more generally, one should continue to search 
for any sign of new physics or new particles, new gauge bosons and fermions, as 
predicted in most of the SM extensions.  

In conclusion, it is not yet time to give up on SUSY and more generally on new physics
but, rather, to work harder to be fully prepared for the more precise and larger data 
set that will be delivered by the upgraded LHC. It will be soon enough to ``philosophize" then 
as the physics landscape will become more clear.\smallskip

\noindent {\bf Acknowledgements:} I would like the thank the organisers for their 
invitation to the school and their kind hospitality. This work is supported by the ERC Advanced Grant Higgs@LHC.

\end{document}